\documentclass[useAMS,usenatbib]{mn2e}

\usepackage{latexsym}
\usepackage{amssymb}
\usepackage{graphicx}

\title[Self-Similar Force-Free Wind From an Accretion
  Disk]{Self-Similar Force-Free Wind From an Accretion Disk}

\author[Ramesh Narayan, Jonathan C. McKinney, and Alison
  J. Farmer]{Ramesh Narayan$^{1}$\thanks{E-mail:
  narayan@cfa.harvard.edu (RN) ;
  jmckinney@cfa.harvard.edu (JCM);
  afarmer@cfa.harvard.edu (AJF)},Jonathan  C. McKinney$^{1}$\footnotemark[1], and Alison J. Farmer$^{1}$\footnotemark[1]\\
  $^{1}$Institute for Theory and Computation, Center for
  Astrophysics, Harvard University, 60 Garden St., Cambridge, MA,
  02138
}

\newcommand{\detg}{{\sqrt{-g}}}

\def\fps@figure{bp}%
\def\fps@table{bp}%
\def\fps@plate{bp}%
\def\eps@scaling{1.0}%
\newcommand\epsscale[1]{\gdef\eps@scaling{#1}}%
\newcommand\plotone[1]{%
 \centering
 \leavevmode
 \includegraphics[width={\eps@scaling\columnwidth}]{#1}%
}%
\newcommand\plottwo[2]{%
 \centering
 \leavevmode
 \columnwidth=.45\columnwidth
 \includegraphics[width={\eps@scaling\columnwidth}]{#1}%
 \hfil
 \includegraphics[width={\eps@scaling\columnwidth}]{#2}%
}%
\newcommand\plotfiddle[7]{%
 \centering
 \leavevmode
 \vbox\@to#2{\rule{\z@}{#2}}%
 \includegraphics[%
  scale=#4,
  angle=#3,
  origin=c
 ]{#1}%
}%

\newcommand\apj{\rmfamily{ApJ}}%
\newcommand\apjl{\rmfamily{ApJ}}%
\newcommand\apss{\rmfamily{Ap\&SS}}%
\newcommand\aap{\rmfamily{A\&A}}%
\newcommand\mnras{\rmfamily{MNRAS}}%
\newcommand\prd{\rmfamily{Phys.~Rev.~D}}%
\newcommand\pasj{\rmfamily{PASJ}}%
\newcommand\nat{\rmfamily{Nature}}%
\let\la=\lesssim            
\let\ga=\gtrsim

\def\vB{\vec{B}}
\def\vE{\vec{E}}
\def\vj{\vec{j}}
\def\vnab{\vec\nabla}
\def\pa{\partial}
\def\Rfp{R_{\rm fp}}
\def\RLC{R_{\rm A}}

\begin{document}

 \date{Accepted 2006 Oct 26. Received 2006 Sept 16; in original form 2006 Sept 16}
  \pagerange{\pageref{firstpage}--\pageref{lastpage}} \pubyear{2006}
  \maketitle \label{firstpage}

\begin{abstract}

We consider a self-similar force-free wind flowing out of an
infinitely thin disk located in the equatorial plane.  On the disk
plane, we assume that the magnetic stream function $P$ scales as
$P\propto R^\nu$, where $R$ is the cylindrical radius.  We also assume
that the azimuthal velocity in the disk is constant: $v_\phi = Mc$,
where $M<1$ is a constant.  For each choice of the parameters $\nu$
and $M$, we find an infinite number of solutions that are physically
well-behaved and have fluid velocity $\leq c$ throughout the domain of
interest.  Among these solutions, we show via physical arguments and
time-dependent numerical simulations that the minimum-torque solution,
i.e., the solution with the smallest amount of toroidal field, is the
one picked by a real system.  For $\nu \geq 1$, the Lorentz factor of
the outflow increases along a field line as $\gamma \approx
M(z/\Rfp)^{(2-\nu)/2} \approx R/R_{\rm A}$, where $\Rfp$ is the radius
of the foot-point of the field line on the disk and $R_{\rm A}=\Rfp/M$
is the cylindrical radius at which the field line crosses the Alfven
surface or the light cylinder.  For $\nu < 1$, the Lorentz factor
follows the same scaling for $z/\Rfp < M^{-1/(1-\nu)}$, but at larger
distances it grows more slowly: $\gamma \approx (z/\Rfp)^{\nu/2}$. For
either regime of $\nu$, the dependence of $\gamma$ on $M$ shows that
the rotation of the disk plays a strong role in jet acceleration. On
the other hand, the poloidal shape of a field line is given by $z/\Rfp
\approx (R/\Rfp)^{2/(2-\nu)}$ and is independent of $M$.  Thus
rotation has neither a collimating nor a decollimating effect on field
lines, suggesting that relativistic astrophysical jets are not
collimated by the rotational winding up of the magnetic field.

\end{abstract}


\normalsize
\section{Introduction}

Although relativistic jets from accreting black holes have been known
and have been studied for decades, the physics behind the production,
acceleration and collimation of these jets is still poorly understood.
Broadly, there are two schools of thought on the launching of jets.
According to the first (e.g., Lovelace 1976), jets are the innermost
and most energetic region of an extended outflow from the black hole
accretion disk.  The power for the jet thus comes from the disk.  In
the other view, the jet is powered by the free energy associated with
the spin of the central black hole (Blandford \& Znajek 1977).  The
jet phenomenon is then a remarkable manifestation of the Penrose
(1969) process.

The fact that jets from black holes are observed to accelerate to
relativistic velocities, despite the absence of any significant
radiative or thermal driving, strongly suggests that magnetic fields
play a dynamically important role.  This motivates the study of
magnetohydrodynamic (MHD) models of jets.  Moreover, given the
difficulty of achieving self-collimation in an isolated jet --- e.g.,
Michel's (1973a) solution for a spinning monopole has no collimation
at all (see \S~3.2 below) --- it is reasonable to suppose that the
presence of an extended outflow surrounding the jet is needed for
maintaining both the collimation and dynamical stability of the jet,
as stressed for instance by Appl \& Camenzind (1992, 1993) and Beskin
\& Malyshkin (2000).  This external flow is presumably a wind of some
sort from the disk.  It is thus of interest to study the dynamics of
extended, nearly self-similar, MHD, disk outflows.

Given the complexity of the full relativistic MHD equations, a variety
of simplifications have been introduced by different authors.  Many
studies make a nonrelativistic approximation (e.g., Blandford \& Payne
1982; Heyvaerts \& Norman 1989; Contopoulos \& Lovelace 1994;
Contopoulos 1995b; Ostriker 1997).  Although it is adequate for
certain aspects of the problem, this approach fails to capture the
presence of a light cylinder (or relativistic Alfven surface), an
explicitly relativistic phenomenon that likely plays an important role
in jet physics.  The reader is referred to the seminal work on light
cylinders by Goldreich \& Julian (1969) who discussed pulsar
magnetospheres and Okamoto (1974, 1978) who considered disk winds, as
well as the work of Lovelace et al. (1986), Nitta, Takahashi \&
Tomimatsu (1991), Beskin \& Pariev (1993) and Beskin (1997) who
included the effects of general relativity.

In considering relativistic winds, one could either work with the full
MHD equations (e.g., Vlahakis \& Konigl 2003), or make further
simplifications.  One possibility is to assume that the outflowing gas
is cold (e.g., Okamoto 1978; Li et al. 1992; Contopoulos 1994) so that
only the inertia of the fluid contributes to the dynamics and there
are no effects due to pressure or internal energy.  An even more
drastic simplification is to ignore gas inertia as well.  In this {\it
force-free} approximation, the plasma supplies charges and currents as
needed to support the electromagnetic fields, but it has no other
dynamical role.

Force-free electrodynamics is the simplest and cleanest model of
magnetized relativistic outflows.  If magnetic fields are at all
important in disk winds and jets, one might hope that the force-free
approximation would capture a good fraction of the relevant physics.
Early discussions of force-free disk winds may be found in Okamoto
(1974), Blandford (1976) and Blandford \& Znajek (1977).  Recently,
force-free electrodynamics has been formulated as a system of
time-dependent equations \citep{kom02b,kom04,mckinney2006a,spit06}.
These time-dependent force-free equations have been used to study
black hole and neutron star magnetospheres
\citep{kom01,kom02a,kom02b,kom04,kom06,mckinney2006a,mckinney2006b,mn06b,spit06}.

Force-free models have gained new relevance as a result of
time-dependent general-relativistic MHD simulations of black hole
accretion flows (e.g., McKinney \& Gammie 2004; Gammie et al. 2004; de
Villiers et al.  2005).  These simulations reveal a nearly force-free
highly relativistic jet forming spontaneously along the symmetry axis
and accelerating rapidly outwards.  Moreover, the height-integrated
azimuthal current within the turbulent disk is found to have a
surprisingly simple power-law behavior (McKinney \& Narayan 2006a,b),
$dI_\phi /dR \propto R^{-5/4}$, where $R$ is the radius and
$I_\phi(R)$ is the azimuthal current enclosed inside $R$.  In
addition, an idealized force-free model with an equatorial rotating
current sheet with azimuthal current varying as $R^{-5/4}$, which is
equivalent to the magnetic stream function varying as $P \propto
R^{3/4}$ (or $\nu=3/4$ in eq. \ref{Peq} below), gives a surprisingly
good match to a variety of properties of the simulated jet such as the
poloidal configuration of the magnetic field, the collimation angle of
the jet, the Lorentz factor, etc. Thus, it appears that simple disk
wind models with self-similar (power-law) scalings may capture key
aspects of the jet problem.  We should note, however, that
self-similar disk wind models with the particular scaling $P \propto
R^{3/4}$ have a checkered history and there has been some confusion on
whether or not well-behaved solutions that extend to large distances
from the disk are present at all.  Blandford \& Payne (1982) and
Contopoulos (1995a), for instance, both find that such solutions do
not exist.

In this paper we analyze the structure of self-similar force-free disk
winds for a range of power-law profiles of the stream function,
including the case $P \propto R^{3/4}$.  \S~2 introduces our model,
which is closely similar to the model of Contopoulos (1995a) and is a
self-similar specialization of Okamoto's (1974) more general analysis.
We show that the problem is reduced to solving the second-order
differential equation (\ref{Tequation}) with appropriate boundary
conditions.  \S~3 discusses a number of analytical results and maps
out the different kinds of solutions that are possible.  \S~4 presents
results from a numerical analysis of the basic differential equation
and identifies regions in parameter space where the different solution
types are found.  \S~5 presents time-dependent numerical simulations
of force-free disk winds using a relativistic force-free code and
draws some general conclusions as to which of the many solutions are
relevant for given situations.  The paper concludes with a discussion
in \S~6.  The Appendix discusses some mathematical properties of the
differential equation (\ref{Tequation}).

\section{The Model}

\subsection{Okamoto's (1974) Force-Free Field Equation}

We consider a steady, axisymmetric force-free field in cylindrical
coordinates: $R, ~\phi, ~z$.  Since $\vnab\cdot\vB=0$, the magnetic
field $\vB$ can be written in terms of a vector potential: $\vB =
\vnab\times \vec{A}$.  Furthermore, since we have assumed axisymmetry
($\pa/\pa\phi=0$), the poloidal component $\vec{B_p}$ of the magnetic
field (the projection on the $Rz$ plane) can depend only on the $\phi$
component $A_\phi$ of the vector potential.  Thus, we may write the
field quite generally as
\begin{eqnarray}
\vB & = & \vec{B_p} +  B_\phi \hat\phi = \vnab\times(A_\phi \hat\phi)
+ B_\phi \hat\phi \nonumber\\
& = & \left[-{1\over R}{\pa(RA_\phi)\over\pa{z}}, ~B_\phi, ~{1\over
R}{\pa(RA_\phi)\over\pa{R}}\right],
\end{eqnarray}
where $B_\phi$ is the toroidal component of the field.  It is standard
to rewrite this in terms of the stream function $P\equiv RA_\phi$ (the
stream function is also often represented by the symbol $\psi$):
\begin{equation}
\vB = \left[-{1\over R}{\pa{P}\over\pa{z}}, ~B_\phi, ~{1\over
R}{\pa{P}\over\pa{R}}\right]. \label{B}
\end{equation}
The magnetic flux enclosed within radius $R$ is given by
\begin{equation}
\Phi(R) = \int_0^R B_z 2\pi R'dR' = 2\pi P(R).
\end{equation}
Thus, apart from a factor of $2\pi$, the stream function $P$ is the
same as the enclosed magnetic flux.

Since the enclosed flux interior to a given field line is constant,
independent of the $z$ at which it is measured, it is clear that $P$
must be a constant along each field line.  This can also be shown
explicitly from equation (\ref{B}) by verifying that $\vnab
P\cdot\vB=0$.  Thus, each field line is uniquely specified by its
value of $P$.  Another quantity that is constant on each field line is
the angular velocity $\Omega$.  Since we imagine that the field line
is embedded in an accretion disk and corotates with it at its
foot-point, the angular velocity of the entire field line out to
infinity is determined by the rotation of the disk at the foot-point.
$\Omega$ is clearly a function only of $P$.

By definition, in the frame rotating with the angular velocity
$\Omega$ of a field line we are in the comoving frame of the fluid and
hence there is no electric field (because we assume infinite
conductivity).  Therefore, back in the nonrotating frame, there must
be an electric field, which is given by
\begin{equation}
\vE = -{\Omega R\over c}\hat\phi\times\vB ={\Omega R\over c} (-B_z,
~0, ~B_R) =-{\Omega\over c}\vnab{P}. \label{E}
\end{equation}
The electric charge density may be obtained
from Gauss's law:
\begin{eqnarray}
\rho_e &=& {1\over 4\pi}\vnab\cdot\vE \\
& = & -{\Omega\over4\pi c}\left({\pa^2P\over\pa R^2} + {1\over R}{\pa
P \over \pa R} + {\pa^2P\over\pa z^2}\right) -{1\over4\pi
c}|\nabla{P}|^2{d\Omega\over dP}\nonumber.
\end{eqnarray}

In steady state the current density is given by
\begin{equation}
\vj \equiv \vec{j_p} + j_\phi \hat\phi = {c\over4\pi}\vnab\times\vB,
\end{equation}
where we have once again decomposed the vector into poloidal and
toroidal components.  This gives
\begin{eqnarray}
j_\phi &=& {c\over4\pi}(\vnab\times\vB)_\phi = -{c\over4\pi
R}\left({\pa^2P\over\pa R^2} -{1\over R}
{\pa P\over\pa R} + {\pa^2P\over\pa z^2}\right), \\
\vec{j_p} &=& {c\over4\pi}(\vnab\times\vB)_p = -{c\over4\pi}{\pa
B_\phi\over\pa z}\hat{R} +{c\over4\pi R}{\pa (RB_\phi)\over\pa
R}\hat{z}. \label{jp}
\end{eqnarray}

In the force-free limit we have no inertial terms in the momentum
equation, so the sum of all the electromagnetic forces at each point
must be zero.  Thus the equation of motion takes the simple form
\begin{equation}
\rho_e\vE + {1\over c} \vj\times\vB = 0. \label{eom}
\end{equation}
Since $\vE$ has no toroidal component, neither can $\vj\times\vB$.
Thus $\vec{j_p}$ and $\vec{B_p}$ have to be parallel to each other.
Comparing the expression for $\vec{j_p}$ in equation (\ref{jp}) with
that for $\vB$ in equation (\ref{B}), we see that the quantity
$RB_\phi$ must be a unique function of $P$.  Hence this is a third
quantity that is constant along a field line:
\begin{eqnarray}
RB_\phi &=& \beta(P), \\
\vec{j_p} &=& {c\over4\pi} {d\beta\over dP} \vec{B_p}. \label{vecjp}
\end{eqnarray}
Note that, at each point, $\beta$ is proportional to the net enclosed
current in the $z$-direction

Taking the dot product of equation (\ref{eom}) with $\vE$, we obtain
\begin{equation}
\rho_eE^2 + {1\over c} \vE\cdot(\vj\times\vB).
\end{equation}
Substituting the expressions written down earlier for the various
quantities, and after some algebra, we finally obtain the following
partial differential equation for $P$:
\begin{eqnarray}
0 &=& \left(1-{\Omega^2R^2\over c^2}\right){\pa^2P\over\pa R^2} -
\left(1+{\Omega^2R^2\over c^2}\right){1\over R}{\pa P\over\pa R} \nonumber\\
&+&\left(1-{\Omega^2R^2\over c^2}\right){\pa^2P\over\pa z^2}
\beta{d\beta\over dP} -{\Omega R^2\over c^2}{d\Omega \over dP} |\nabla
P|^2 . \label{okamoto}
\end{eqnarray}
This is the force-free field equation of Okamoto (1974, compare with
his eq. 43; simpler versions of this equation without the last term
were discussed earlier by Mestel 1973, Michel 1973b and Scharlemann \&
Wagoner 1973).

\subsection{Self-Similar Disk Wind}

The discussion so far was quite general.  All we assumed were the
following conditions: (i) steady state, (ii) axisymmetry, (iii)
force-free (i.e., negligible inertia), (iv) infinite conductivity. Now
we explicitly consider the problem at hand, viz., a {\it self-similar
force-free wind} flowing out of an accretion disk (Contopoulos 1995a).
We assume that the disk is infinitely thin and located at the
equatorial plane, $z=0$.  Each field line is identified by the radius
$\Rfp$ of its foot-point on the disk.  The disk supplies two boundary
conditions at the foot-point:

\noindent 1. The magnetic flux $\Phi(\Rfp)$ enclosed inside radius
$\Rfp$: This determines the stream function $P(\Rfp)=\Phi(\Rfp)/2\pi$
of each field line as a function of the radius of its footpoint.

\noindent 2. The angular velocity $\Omega(\Rfp)$ of the disk at radius
$\Rfp$: The foot-point of the field line is dragged around by the disk
at the angular velocity $\Omega(\Rfp)$ and since $\Omega$ is constant
along each field line this angular velocity is imparted to the entire
line out to infinity.

Assuming that we have a self-similar disk with a power-law structure,
we write the stream function on the disk plane as
\begin{equation}
P(R,z) \big|_{z=0} \propto R^\nu, \qquad 0 \leq \nu \leq 2.
\label{Peq}
\end{equation}
By this definition, $\nu$ is equivalent to the index $x$ in
Contopoulos (1995a) and the index $F$ in Vlahakis \& Konigl (2003).
The vertical component of the magnetic field on the disk plane then
scales as
\begin{equation}
B_z(R,z) \big|_{z=0} = \left({1\over R}{\pa P\over \pa R}\right)_{z=0}
\propto R^{\nu-2}.
\end{equation}
Since we would like the enclosed magnetic flux to vanish as $R\to0$,
we impose the condition $\nu\geq0$.  Also, since we are not interested
in solutions in which the magnetic field strength increases with
increasing radius, we restrict ourselves to $\nu \leq 2$.  Note that
the monopole solution of Michel (1973a) corresponds to $\nu=0$, the
problem considered by Blandford \& Payne (1982) corresponds to
$\nu=3/4$, the paraboloidal field model of Blandford (1976)
corresponds to $\nu=1$, and the models discussed by Ostriker (1997)
correspond to the range $1 < \nu < 2$.

The power-law form of the boundary condition (\ref{Peq}) motivates us
to consider the following self-similar form for the stream function,
\begin{equation}
P(R,z) = R^\nu T(u), \label{Pselfsim}
\end{equation}
where the self-similar coordinate $u$ (called $Z$ by Contopoulos
1995a) is given by
\begin{equation}
u\equiv {z\over R}.
\end{equation}
For convenience we set
\begin{equation}
T(0) = 1, \label{bc1}
\end{equation}
which is equivalent to absorbing any coefficient on the right-hand
side of equation (\ref{Pselfsim}) into the definition of $R$.  Since
field lines live on surfaces of constant $P$, a field line going
through the point $(R,z)$ has its foot-point at radius
\begin{equation}
\Rfp = R T^{1/\nu}(z/R).
\end{equation}
Equivalently, for a given $\Rfp$, the shape of the field line in the
poloidal plane is given by the parametric equations
\begin{equation}
R(u) = \Rfp T^{-1/\nu}(u), \qquad  z(u) = \Rfp uT^{-1/\nu}(u).
\label{fieldline}
\end{equation}

Consider now the angular velocity.  From equation (\ref{okamoto}), it
is clear that the dimensionless ratio $\Omega R/c$ plays a prominent
role in this model.  In order to obtain a self-similar solution, this
quantity has to be constant on the disk plane.  Thus we write
\begin{equation}
\left({\Omega R\over c}\right)_{z=0} = M = {\rm constant},
\label{Omeq}
\end{equation}
where $M$ ($=1/c_n$ in Contopoulos 1995a) represents a sort of ``Mach
number'' of the angular motion.  Note that the azimuthal velocity
follows a ``flat rotation curve,'' not a Keplerian profile.  This is
not the ideal choice for an accretion disk, but it is the only profile
that is consistent with self-similarity (e.g., Li et al. 1992;
Contopoulos 1995a; Vlahakis \& Konigl 2003)\footnote{This restriction
arises because we are working with a relativistic theory in which $c$
introduces a fundamental unit of velocity.  Non-relativistic
self-similar models have the freedom to choose any power-law for the
radial profile of the angular velocity, including in particular a
Keplerian profile, but the price for this is that they ignore all
relativistic effects.}.  Since $\Omega$ is constant on a field line,
the angular velocity at any general $(R,z)$ is given by
\begin{equation}
{\Omega R\over c} = M T^{-1/\nu}. \label{Omega}
\end{equation}

Consider now the third conserved quantity on a field line, the
quantity $\beta$.  Under self-similarity, we expect the ratio of the
toroidal and poloidal field strengths at the disk plane to be a
constant.  Thus we assume that
\begin{equation}
\left({B_\phi \over B_z}\right)_{z=0} = -sM = {\rm constant}.
\label{sdef}
\end{equation}
We include $M$ on the right-hand side since we expect the amount of
toroidal field in the wind to be roughly proportional to the rotation
rate of the disk.  We also introduce a negative sign because we expect
field lines to be swept {\it back} with respect to the direction of
rotation.  (The quantity $H_0$ in Contopoulos 1995a is $-sM$ in our
notation.)  We then obtain
\begin{eqnarray}
\beta &=& -\nu sM R^{\nu-1}T^{(\nu-1)/\nu}, \\
\beta{d\beta\over dP} &=& \nu (\nu-1)s^2M^2 R^{\nu-2}T^{(\nu-2)/\nu}.
\end{eqnarray}

Substituting the various self-similar scalings written down above into
equation (\ref{okamoto}), we finally obtain the differential equation
satisfied by the similarity function $T(u)$:
\begin{eqnarray}
0 &=& (u^2+1)T'' + (3-2\nu)u T' + \nu(\nu-2) T \nonumber\\
&-& M^2T^{-(\nu+2)/\nu} \bigg{[}(u^2+1)TT'' -(u^2+1){T'}^2/\nu \nonumber\\
&~& \qquad  +(3-2\nu)uTT' + \nu(\nu-1)(1-s^2)T^2 \bigg{]} .
\label{Tequation}
\end{eqnarray}
This is the fundamental equation that we need to solve to study
self-similar force-free disk winds.  Once we have a solution for
$T(u)$, we can calculate all other quantities of interest.  For
instance, the three components of the magnetic field are given by
\begin{eqnarray}
B_R(R,z) &=& -R^{\nu-2} T'(u), \label{BR}\\
B_\phi(R,z) &=& -\nu sM R^{\nu-2} T^{(\nu-1)/\nu}(u), \label {Bphi}\\
B_z(R,z) &=& R^{\nu-2} [\nu T(u) - u T'(u)], \label{Bz}
\end{eqnarray}
where as always $u=z/R$.

\subsection{The Alfven Critical Surface}

The highest derivative $T''$ in equation (\ref{Tequation}) is
multiplied by a factor of the form
\begin{equation}
\left(1-M^2T^{-2/\nu}\right) \equiv \left(1-{\Omega^2R^2\over
c^2}\right), \label{LC}
\end{equation}
and so the differential equation is clearly singular when this factor
goes to zero.  The singularity corresponds to the Alfven critical
surface, also called the light cylinder in some of the earlier
literature.  It is located at the radius
\begin{equation}
\RLC = {\Rfp\over M}.
\end{equation}
Note that the Alfven surface occurs at a specific value of $u=u_c$ and
has the shape of a cone.  This is in contrast to the pulsar
magnetosphere problem where it occurs at a single radius, thereby
motivating the terminology ``light cylinder'' in that case.

When $u=u_c$ the factor given in eq (\ref{LC}) goes to zero, and
therefore in order for the solution to be well-behaved the rest of the
terms in (\ref{Tequation}) that do not involve $T''$ should also add
up to zero.  We thus obtain the following two regularity conditions at
$u_c$:
\begin{eqnarray}
T(u_c) &=& M^\nu, \label{bc2}\\
T'(u_c) &=& -\nu M^\nu \left[{1-(\nu-1)s^2 \over u_c^2+1}
\right]^{1/2}. \label{bc3}
\end{eqnarray}
The negative sign on the second condition is required for a physically
sensible solution.

For a given solution $T(u)$, it is easy to determine whether any local
stretch of the solution is inside or outside the Alfven surface ; here
and elsewhere, we use the term ``inside'' for $u<u_c$ and ``outside''
for $u>u_c$. If $T(u) > T(u_c) = M^\nu$, then we are inside the Alfven
surface, and if $T(u) < M^\nu$ we are outside.  Note that, for
$\nu<1$, the condition (\ref{bc3}) can be satisfied for any value of
$s$.  However, for $\nu>1$, a solution is possible only if $s^2 \leq
1/(\nu-1)$.  If $|s|$ is larger than this limit, the solution cannot
pass through the Alfven surface.

\subsection{Flow Velocity}
\label{velocity}

While the azimuthal velocity of a field line passing through $(R,z)$
is given by equation (\ref{Omega}), this does not determine the {\it
fluid velocity}.  By assumption, the plasma has no inertia, so we are
allowed to add any velocity parallel (or antiparallel) to the field
line without changing the solution in any way.  There is thus
considerable ambiguity as to what we mean by the fluid velocity.  We
could, however, ask what the {\rm minimum} fluid velocity is, i.e., we
could choose the parallel velocity such that the net velocity of the
plasma is a minimum.  This minimum velocity is the ``particle drift
velocity,''
\begin{equation}
v_{\rm min} \equiv {|\vec{S}| \over B^2/4\pi},
\end{equation}
where
\begin{eqnarray}
\vec{S} &=& {c\over4\pi}\vec{E}\times\vec{B} \nonumber\\
&=& {\Omega R\over4\pi} \left[-B_\phi B_R\hat{R}
+(B_R^2+B_z^2)\hat{\phi} - B_\phi B_z\hat{z}\right]
\end{eqnarray}
is the Poynting flux, $\vec{E}$ is the electric field given in
equation (\ref{E}), and $B$ is the total magnetic field strength. We
then have
\begin{equation}
{v_{\rm min} \over c} = \Omega R {B_p\over B} = {R\over R_{\rm
LC}}{B_p\over B} = M{R\over R_{\rm fp}} {B_p \over B}, \label{vmin}
\end{equation}
where $B_p$ is the magnitude of the poloidal component of the magnetic
field.

Another relevant speed is the ``energy flow speed,'' which is given by
\begin{equation}
v_E \equiv {|\vec{S}|\over U} ,
\end{equation}
where $U$ is the electromagnetic energy density given by
\begin{equation}
U={1\over8\pi}\left[|\vec{E}|^2+|\vec{B}|^2\right] =
{B_p^2\over8\pi}\left({\Omega^2R^2\over c^2}+1\right)
+{B_\phi^2\over8\pi}.
\end{equation}
The energy flow speed is related to the particle drift speed by
\begin{equation}
v_E = {2v_{\rm min} \over 1 + v_{\rm min}^2/c^2},
\end{equation}
which shows that $v_E$ is always $\leq c$.  Intuitively, one expects a
physically sensible solution to have $v_E,v_{\rm min} \to c$ at large
distance from the disk, which is equivalent to the condition that the
fast critical surface is at infinity. This requirement is met when the
following two conditions are satisfied:
\begin{equation}
-{B_\phi\over B_p} \to {\Omega R\over c} \to \infty.
\end{equation}
The second condition says that we have to be infinitely far outside
the Alfven surface, i.e., we should focus on the ``paraboloidal''
solutions discussed in \S~3.

So long as we are inside the Alfven surface, we are guaranteed that
$v_{\rm min} <c$.  However, once a field line goes outside the Alfven
surface ($\Omega R > c$), there is nothing in the problem to prevent
$v_{\rm min}$ from exceeding $c$.  This highlights a major weakness of
the force-free model.  Even though the model is fully relativistic and
causal, it has no way of enforcing $v<c$ for the fluid it is meant to
describe because it does not explicitly make use of the inertia. Note
that in any region that has $v_{\rm min} > c$, the electric field
strength exceeds the magnetic field strength, allowing unlimited
electrostatic acceleration of charged particles.

In this paper, we focus our attention on solutions that have $v_{\rm
min} \leq c$ for all $u$ of interest, i.e., all $R$ and $z$ spanned by
a given problem.  We call such solutions physically viable and
consider solutions that violate this condition as ``unphysical.''  The
restriction to physical solutions plays the role of a boundary
condition on the solution (see \S~\ref{secbc}).  From equation
(\ref{vmin}) we see that requiring the solution to be physical is
equivalent to requiring a minimum amount of toroidal field, or
equivalently, a minimum level of enclosed $j_z$ (see the comment below
eq. \ref{vecjp}).

We should note, however, that even solutions we consider unphysical
under the above strict criterion may have a role under some
circumstances.  For instance, an MHD flow with plasma inertia may have
a force-free zone that matches the corresponding segment of one of our
unphysical solutions but may then deviate into a non-force-free flow
in the region where the unphysical solution has $v_{\rm min} > c$
(e.g., Contopoulos 1995a).  A comparison of the force-free solutions
discussed in this paper with self-similar MHD solutions would thus be
very instructive.

\subsection{Boundary Conditions}
\label{secbc}

Let us count the number of degrees of freedom in the problem and the
boundary conditions we need.  Since equation (\ref{Tequation}) is a
second order differential equation, it requires two boundary
conditions.  The constants $\nu$ and $M$ are clearly determined by the
properties of the disk and should be treated as externally supplied
parameters of the problem.  However, the other two constants, $s$ and
$u_c$, are not free parameters.  They have to be determined
self-consistently in the process of solving the problem, i.e., they
are {\it eigenvalues}.

Consider first a problem in which field lines cross the Alfven
surface. To determine the two eigenvalues $s$ and $u_c$ we have the
two regularity conditions (\ref{bc2}) and (\ref{bc3}).  Of the two
boundary conditions needed by the differential equation, we have
already discussed one, viz., the trivial condition (\ref{bc1}).  The
second boundary condition must be set at infinity.  For this purpose
we make use of the velocity constraint $v_{\rm min} \leq c$ discussed
in \S~\ref{velocity}, except that we now require $v$ to be exactly
equal to $c$ at infinity:
\begin{equation}
\left({v_{\rm min}\over c}\right)_{u\to\infty}= \left(M {R\over\Rfp}
{B_p\over B}\right)_{u\to\infty} =1. \label{bc4}
\end{equation}
This condition is equivalent to the statement that we pick the
physically allowed solution with the minimum level of toroidal field.
In the language of Michel (1969), it is the ``minimum torque'' (or
equivalently, minimum energy) solution.  Such solutions have the
minimum angular momentum and energy flux that connect to asymptotic
infinity (Kennel et al. 1983).

The logic for picking this particular solution will become clearer in
\S~3.3.3 below.  The condition (\ref{bc4}) states that the ``fast
critical surface'' or the ``light surface,'' i.e., the point at which
the fluid speed is equal to $c$, is located at infinity, and not
``beyond infinity'' (see Okamoto 1974, who calls this surface the
Alfven point).  If we are interested in a system that extends to a
finite maximum value of $u=u_{\rm max}$ rather than to infinity, we
may wish to use the same condition (\ref{bc4}) but applied at $u_{\rm
max}$.  This would be equivalent to an outflowing boundary condition
on the force-free flow.

When we consider a problem in which field lines are entirely inside
the Alfven surface, we need two boundary conditions plus a third
condition to determine the eigenvalue $s$.  In this case, we have only
one boundary condition (eq.  \ref{bc1}), and the other two conditions
must be set at infinity.  Given a particular solution such as those
discussed in \S~3.5, it is straightforward to come up with the
necessary boundary conditions.

\section{Analytical Results}

\subsection{Blandford's Paraboloidal Solution}

When $\nu=1$, the differential equation (\ref{Tequation}) has an
analytical solution:
\begin{eqnarray}
T(u) &=& (1+u^2)^{1/2}-u , \nonumber\\
P(R,z) &=& (R^2+z^2)^{1/2} - z . \label{para}
\end{eqnarray}
This is the self-similar version of the paraboloidal solution of
Blandford (1976).  The location $u_c$ of the Alfven surface is obtained by solving
the condition (\ref{bc2}), which gives
\begin{equation}
u_c = {(1-M^2) \over 2M}.
\end{equation}
The components of the magnetic field are given by
\begin{eqnarray}
B_R(R,z) &=& {1\over R} - {z\over R(R^2+z^2)^{1/2}}, \label{parBR} \\
B_\phi(R,z) &=& -{sM \over R}, \label{parBphi} \\
B_z(R,z) &=& {1\over (R^2+z^2)^{1/2}}. \label{parBz}
\end{eqnarray}
The poloidal projection of a field line with footpoint at $R_{\rm fp}$
has the shape
\begin{equation}
z = {1\over 2R_{\rm fp}} \left(R^2 - R_{\rm fp}^2\right),
\end{equation}
which shows explicitly the parabolic geometry of the field (hence the
name of this solution).

Note that the poloidal field structure is independent of the rotation
parameter $M$ as well as the eigenvalue $s$.  In order to determine
$s$, we must apply the boundary condition (\ref{bc4}). Substituting
equations (\ref{parBR}--\ref{parBz}) in equation (\ref{vmin}), we find
\begin{equation}
\left({v_{\rm min} \over c}\right)_{u\to\infty} = {2 \over |s|}.
\label{paravmin}
\end{equation}
Thus the boundary condition (\ref{bc4}) gives
\begin{equation}
s = 2.
\end{equation}
Only for this value of $s$ does the fluid velocity $v_{\rm min}$
asymptotically tend to $c$ at infinity.  ($s=-2$ is also a solution,
but this corresponds to a Poynting flux pointed towards the disk and
is not of interest.)  For $s < 2$, $v_{\rm min}$ exceeds $c$ beyond a
certain value of $u$, and the solutions are unphysical.  For $s > 2$,
$v_{\rm min}$ is less than or equal to $c$ as $u\to\infty$ and it
would appear that all the solutions with $s>2$ are physical.  Within
the framework of the pure self-similar problem, there is nothing to
indicate that any one of these solutions is more physical than the
others.  However, the more complete analysis given by Blandford
(1976), which includes a regularity condition as $R\to0$, shows that
only $s=2$ is consistent with a non-singular solution on the axis.
All other values of $s$ that appear to be physically valid within the
self-similar analysis are, in fact, unphysical since they require a
singular current along the axis.  This point will become clearer in
\S~3.3.3.

The fact that $B_R$ and $B_z$ of the paraboloidal solution are
independent of $M$ means that the poloidal structure is independent of
rotation.  Thus, the presence of the azimuthal field $B_\phi$, however
strong it might be, has no collimating or decollimating effect.
Roughly, one could say that the collimating hoop stress associated
with the toroidal field is exactly canceled by the decollimating
effect of the pressure gradient associated with the same field (see
Ostriker 1997).

For the solution with $s=2$, we can calculate the Lorentz factor
$\gamma_{\rm min}$ corresponding to the minimum fluid velocity $v_{\rm
min}$.  Assuming $u \gg 1$, we find as a function of distance from the
disk plane
\begin{equation}
\gamma_{\rm min} \approx {2M \over (1-M^2)^{1/2}} \left( {z \over
\Rfp} \right)^{1/2} \approx \sqrt{2} \left[{R^2-\Rfp^2 \over R_{\rm
LC}^2-\Rfp^2}\right]^{1/2}, \label{pargamma}
\end{equation}
where $\Rfp$ is the radius of the foot-point of the field line under
consideration.  Beskin \& Nokhrina (2006) derive and discuss the
asymptotic scaling, $\gamma_{\rm min} \approx R/R_{\rm A}$, for a
relativistic MHD outflow.

\subsection{Michel's Monopole Solution}

When $\nu=0$, the differential equation (\ref{Tequation}) again has an
analytical solution:
\begin{eqnarray}
T(u) &=& 1 - {u\over(1+u^2)^{1/2}} ,\nonumber\\
P(R,z) &=& 1 - {z\over (R^2+z^2)^{1/2}} .
\end{eqnarray}
The components of the magnetic field are given by
\begin{eqnarray}
B_R(R,z) &=& {R\over (R^2+z^2)^{3/2}}, \cr B_\phi(R,z) &=&
-{\Omega\over c} {R\over (R^2+z^2)}, \cr B_z(R,z) &=& {z\over
(R^2+z^2)^{3/2}},
\end{eqnarray}
where, as in the paraboloidal case, we have used the boundary
condition (\ref{bc4}) to determine the coefficient of $B_\phi$.  This
is the monopole solution of Michel (1973a), with the angular velocity
parameter $\Omega$ replacing $M$.

The monopole solution is not really a disk wind solution but rather
corresponds to a wind emitted from a rotating star.  However, by
stitching together copies of the monopole solution with opposite signs
in the two hemispheres it is easy to generate a split monopole
solution.  This has a current sheet in the equatorial plane, just like
the other solutions described in this paper, and is more like a disk
wind.  In both the pure monopole solution and the split monopole
solution, the poloidal field is purely radial and its geometry is
independent of the angular velocity of the star $\Omega$.  Thus, as in
the paraboloidal solution, rotation induces field lines to sweep back
(see the expression for $B_\phi$), but it has no tendency to collimate
the poloidal field.

The Alfven surface is at a fixed radius $R_{\rm A} = \Omega/c$ and
takes the form of a cylinder (the ``light cylinder'').  This is not
the case for any other value of $\nu$.

\subsection{Asymptotic Power-Law Solution at Large $u$ for General $\nu$}

The two analytic solutions discussed above both have $T(u)$ decaying
as a power-law at large $u$; Blandford's paraboloidal solution
($\nu=1$) goes as $T(u) \propto 1/u$ and Michel's monopole solution
($\nu=0$) goes as $T(u) \propto 1/u^2$.  We now consider a general
value of $\nu$ and seek a power-law solution of the form $T(u) \propto
1/u^\mu$ with positive $\mu$.  For convenience, we will refer to such
solutions as ``paraboloidal'' since the field lines have the shape of
generalized parabolae (see eq. \ref{zvsR}).

In the limit of large $u$, we are well outside the Alfven surface
(recall that ``outside'' means $u > u_c$), and the terms proportional
to $M^2$ in equation (\ref{Tequation}) dominate over the other terms.
We may thus neglect the subdominant terms and we may also replace
$u^2+1$ by $u^2$.  We are then left with the simpler equation
\begin{equation}
u^2TT'' -u^2T'^2/\nu + (3-2\nu)uTT' +\nu(\nu-1)(1-s^2)T^2=0,
\label{asympeq}
\end{equation}
which has a power-law solution of the form
\begin{equation}
T(u) \propto u^{-\mu}, \qquad \mu = \nu (s-1), \qquad \nu \neq 0, ~1.
\label{muLC+}
\end{equation}
The above solution for $\mu$ is not valid when $\nu = 0$ or 1 because
the term with $s^2$ in equation (\ref{asympeq}) vanishes for these
cases.  However, we have fully analytic solutions for precisely these
values of $\nu$, as discussed in the previous subsections.  Here we
are interested in other general values of $\nu$.  In order to be
outside the Alfven surface, we require $T(u) < M^\nu$ (as discussed
earlier), and this is asymptotically possible only if $\mu$ is
positive.  Thus we require $s>1$.  Using equation (\ref{fieldline}) we
find that the poloidal shape of the field lines is given by
\begin{equation}
z \propto R^{s/(s-1)}, \qquad R \propto z^{(s-1)/s}. \label{zvsR}
\end{equation}
The shape is vaguely paraboloidal in the sense that $z$ goes as a
power of $R$, though the exponent is not exactly 2.

Given the above solution for $\mu$, the magnetic field components can
be calculated via equations (\ref{BR}--\ref{Bz}) and the fluid
velocity $v_{\rm min}$ can be obtained from equation (\ref{vmin}).  It
is easily verified that any asymptotic power-law solution of the form
(\ref{muLC+}) automatically satisfies the boundary condition $v_{\rm
min} = c$ at $u=\infty$.

Appendix A discusses the properties of these power-law solutions in
some detail.  Equation (\ref{Rexpr}) gives a more accurate estimate of
the shape of the field line, including a first order correction term.
Expressions are also given for the scalings of the three components of
the magnetic field and the minimum Lorentz factor $\gamma_{\rm min}$.
The results for the latter generalize the work of Beskin \& Nokhrina
(2006) who consider the perfect paraboloidal case $\nu=1$; however,
they analyze the MHD problem whereas we consider the simpler
force-free case.

To conclude this subsection, we write down the scalings of various
quantities corresponding to two regimes: $\nu>1$ and $\nu<1$.  These
are the key results of this paper.  The scalings are derived in the
Appendix and are discussed further in \S~4.  We also provide
additional insight into the boundary condition at infinity.

\subsubsection{Scalings for $\nu \geq 1$}

As shown in the Appendix, the case $\nu>1$ is simple.  For any given
$\nu$, there is only one value of $s$ for which there is a
paraboloidal solution; this value is almost exactly equal to $2/\nu$.
The asymptotic form of the solution is
\begin{equation}
T(u) \approx u^{-(2-\nu)},
\end{equation}
and the poloidal shape of a field line is given by
\begin{equation}
{z\over\Rfp} \approx \left({R\over\Rfp}\right)^{2/(2-\nu)}.
\end{equation}
We give only the leading order terms here, and the reader is referred
to Appendix A.2 for the next order terms.  The three components of the
magnetic field scale as follows along a field line:
\begin{eqnarray}
B_R\left({z\over\Rfp}\right) &\approx& \left({\Rfp\over z}\right)^
{(4-\nu)/2}, \\
B_\phi\left({z\over\Rfp}\right) &\approx& -M \left({\Rfp\over
z}\right)^
{(2-\nu)/2}, \\
B_z\left({z\over\Rfp}\right) &\approx& \left({\Rfp\over z}\right)^
{(2-\nu)}.
\end{eqnarray}
The minimum Lorentz factor varies with $z$ as
\begin{equation}
\gamma_{\rm min} \approx M \left({z\over\Rfp}\right)^ {(2-\nu)/2}
\approx M {R\over\Rfp} = {R\over R_{\rm A}}.
\end{equation}
The final expression, $R/R_{\rm A}$, is identical to the result
obtained by Beskin \& Nokhrina (2006).

\subsubsection{Scalings for $\nu < 1$}

When $\nu < 1$, paraboloidal solutions are present for all values of
$s > 1$, and there is no longer a unique solution.  However, the
``correct'' solution is still very close to $s=2/\nu$ (\S\S~4, 5), so
the scalings given above for $T(u)$, $z/\Rfp$, $B_R$, $B_\phi$, $B_z$,
are all approximately valid. The minimum Lorentz factor now has a more
complicated dependence.  We find
\begin{eqnarray}
\gamma_{\rm min} &\approx& M \left({z\over\Rfp}\right)^ {(2-\nu)/2}
\approx {R\over R_{\rm A}}, \qquad {z\over\Rfp} < u_0 ,
\\ \gamma_{\rm min} &\approx& \left({z\over\Rfp}\right)^ {\nu/2},
\qquad {z\over\Rfp} > u_0,
\end{eqnarray}
where $u_0\equiv M^{-1/(1-\nu)}$. We note that the Beskin \& Nokhrina
(2006) scaling $\gamma_{\rm min} \approx R/R_{\rm A}$ is valid for
small values of $z$, but the scaling is different at large distance.
The latter regime may be interpreted as the acceleration produced by
field line curvature, whereas the former regime represents
acceleration as the result of a ``slingshot'' effect (V. Beskin,
private communication).

\subsubsection{Boundary Condition at Infinity as a Regularity Condition on the Axis}

The paraboloidal solutions described above asymptotically have $\Omega
R \gg c$, and their poloidal streamlines are almost precisely
cylindrical.  Let us, therefore, consider a purely cylindrical
force-free problem in which $B_R=0$, and $B_\phi$, $B_z$ and $\Omega$
are functions only of $R$.  The equilibrium condition (Grad-Shafranov
equation) for such a flow takes the simple form,
\begin{equation}
{dB_z^2\over dR} -{1\over R^2}{d\over dR}\left(R^2 {\Omega^2R^2\over
c^2}B_z^2\right) + {1\over R^2}{d\over dR}\left(R^2B_\phi^2\right)=0.
\end{equation}
Asymptotically far from the disk our self-similar solutions have
$B_\phi \gg B_z$, and so the first term is negligibly small; in fact,
for the particular self-similar solution with $s=2/\nu$ discussed in
\S~3.3.1, $B_z$ is independent of $R$ at a fixed asymptotic $z$, so
this term vanishes exactly.  The remaining two terms then give
\begin{equation}
B_\phi^2 = {\Omega^2R^2\over c^2} B_z^2 + C,
\label{Bphicyl}
\end{equation}
where $C$ is an arbitrary integration constant.  For a given $B_z$, we
have an infinite number of solutions, each with a different value of
$C$.  This is equivalent to the freedom we had in choosing $s$ in \S\S
3.1, 3.2.  However, as we now show, only one of these solutions is
well-behaved.

Although we have focused on self-similar solutions in this paper, we
imagine that the solutions are modified inside a ``core'' region at
small radii so as to be analytic on the axis.  The rotation profile,
for instance, might behave as a power law at large $R$ but we expect
it to asymptote to a finite constant value as $R\to0$.  Equation
(\ref{Bphicyl}) then shows that $B_\phi \to C^{1/2}$ on the axis.
However, any solution that has a finite $B_\phi$ as $R\to0$ will have
a singular current along the axis, which might be considered
undesirable.  Therefore, let us impose the additional condition that
$B_\phi\to 0$ as $R\to0$.  This can be satisfied only if $C=0$, and so
we obtain the unique solution
\begin{equation}
{B_\phi\over B_z} = -{\Omega R\over c}. \label{Bphiz}
\end{equation}
By (\ref{Bphicyl}) this relation must be true not only near the axis
but at all $R$, including the self-similar region outside the core.
Thus, analyticity on the axis picks out a unique solution even in the
far-field self-similar zone.  Blandford (1976) reached the same
conclusion for the specific case $\nu=1$.

In the self-similar region, our solutions have $B_\phi \gg B_z$, so
the quantity on the left-hand side of (\ref{Bphiz}) is equal to
$B/B_p$, while that on the right is equal to $R/R_{\rm A}=MR/\Rfp$.
Thus, the condition (\ref{Bphiz}) is the same as
\begin{equation}
M{R\over\Rfp}{B_p\over B} = 1,
\end{equation}
i.e., the same as the condition (\ref{bc4}).  In other words, the
boundary condition $v_{\rm min}\to c$ at infinite distance, which we
previously argued is a physically motivated boundary condition, is in
some sense equivalent to the requirement that the solution be analytic
on the axis.  This provides additional insight on the boundary
condition (\ref{bc4}).

\subsection{Power-Law Solution Inside the Alfven Surface}

Consider next the region well inside the Alfven surface.  Here we may ignore the
terms proportional to $M^2$ in equation (\ref{Tequation}).  Let us
also assume $u^2 \gg 1$.  The differential equation then simplifies to
\begin{equation}
u^2T'' + (3-2\nu)uT' + \nu(\nu-2)T = 0.
\end{equation}
This has a power-law solution
\begin{equation}
T(u) \propto u^{-\mu}, \qquad \mu = -\nu, ~2-\nu. \label{muLC-}
\end{equation}
Since we assume $\nu>0$, the solution $\mu=-\nu$ corresponds to $T(u)$
increasing with $u$.  Asymptotically, this gives a solution with a
poloidal streamline of the form
\begin{equation}
z \propto R^0 = {\rm constant}.
\end{equation}
Thus, as $u\to\infty$, the field line asymptotes to a constant value
of $z$ and the radius $R$ tends to zero.  Thus, field lines converge
radially onto the $z$-axis.  This is clearly unphysical since it
implies a non-zero magnetic monopole density on the axis.
Interestingly, the self-similar solutions described by Blandford \&
Payne (1982) are of this kind and are unphysical.

The second solution in (\ref{muLC-}), $\mu=2-\nu$, has $T(u)$
decreasing with increasing $u$ (so long as $\nu < 2$).  This
asymptotic solution is valid only over the restricted range of $u$
satisfying $u_c \gg u \gg 1$.  For the specific case of a non-rotating
solution, $u_c\to\infty$, and the solution is valid for all $u\gg1$.
The poloidal shape of the field line is given by
\begin{equation}
z \propto R^{2/(2-\nu)}.
\end{equation}
This corresponds to a perfect parabola when $\nu=1$ (Blandford's
solution), and is a generalized parabola for other values of $\nu$.

\subsection{Asymptotic Cylindrical Solution}

Consider next solutions in which field lines asymptotically become
cylindrical with a finite value of $R$.  That is, as $z\to\infty$, the
radius $R$ of each field line tends to a finite value.  Equivalently,
as $u\to\infty$, we have both $T''$ and $T'$ tending to 0 and $T$
tending to a finite value.  Note that these conditions are different
from the quasi-cylindrical asymptotics we discussed in \S~3.3.3.
There, $R/\Rfp$ of a field line went to infinity as $u\to\infty$,
whereas here we consider solutions in which $R/\Rfp$ remains finite.

Given the above conditions, most of the terms in equation
(\ref{Tequation}) vanish and we immediately obtain
\begin{equation}
T(u) \big|_{u\to\infty} = \left[{(\nu-1)(1-s^2) \over
(\nu-2)}\right]^{\nu/2}.
\end{equation}
The asymptotic radius $R_\infty$ of the field line is related to the
radius of the footpoint by
\begin{equation}
{R_\infty \over R_{\rm fp}} = \left[{(\nu-1)(1-s^2) \over
(\nu-2)}\right]^{-1/2},
\end{equation}
and the asymptotic components of the magnetic field are given by
\begin{eqnarray}
B_R &=& 0, \cr B_\phi &=& -R^{\nu-2} \nu sM \left[{(\nu-1)(1-s^2)
\over (\nu-2)}\right]^{1/2}, \cr B_z &=& R^{\nu-2} \nu
\left[{(\nu-1)(1-s^2) \over (\nu-2)}\right].
\end{eqnarray}
The azimuthal velocity of the field line is
\begin{equation}
\left({v_\phi\over c}\right)_{z\to\infty} = \left[{(\nu-1)(1-s^2)
\over (\nu-2)}\right]^{-1/2}.
\end{equation}

The above expressions give real numbers only if the quantity in the
square parentheses is positive.  If $\nu>1$, this requires $s^2 > 1$,
and if $\nu<1$, it requires $s^2 < 1$.  These are necessary conditions
for the existence of cylindrical solutions.

Note that asymptotically cylindrical solutions may be either inside or
outside the Alfven surface.  Whether a particular solution is inside
or outside is determined by whether $R_\infty$ is less than or greater
than $R_{\rm A}$, or equivalently, by whether $v_\phi/c$ is less than
or greater than unity.  If the quantity in square parentheses is less
than 1, then the solution is outside the Alfven surface.  Clearly,
this is always the case when $\nu<1$.  When $\nu>1$, there is no
apparent restriction, though in practice only solutions inside the
Alfven surface are present (see \S~4).

\subsection{Conical Solution}

These are solutions in which $T$ goes to zero at a finite value of $u
= u_f$, so that field lines asymptotically have a conical shape with
$z/R = u_f$ as $z$, $R \to \infty$.  The solution does not exist for
$R < z/u_f$, i.e., inside the cone.  Thus, for force balance, one must
supply the required pressure at the surface of the cone.  Because this
is somewhat unphysical, and also since $v_{\rm min} > c$ as $u \to
u_f$, we do not consider these solutions interesting.  Nevertheless,
we discuss them briefly for completeness.

As stated above, we have $T \to 0$ as $u \to u_f$.  In this limit, the
dominant terms in equation (\ref{Tequation}) are the first and second
terms inside the square brackets.  Focusing on these, the solution
must satisfy
\begin{equation}
\nu T T'' = {T'}^2,
\end{equation}
which has a power-law solution of the form
\begin{equation}
T(u) \propto (u_f-u)^{\nu/(\nu-1)}, \qquad \nu > 1.
\end{equation}
This is the conical solution, which is present only for $\nu>1$.

\section{Numerical Solutions}

Before describing the numerical solutions, let us recall the
properties of Blandford's (1976) paraboloidal solution ($\nu=1$,
\S~3.1).

\begin{description}
\item{(i)} The poloidal structure of field lines is perfectly
parabolic.

\item{(ii)} Rotation has neither a collimating nor a decollimating
effect on the poloidal configuration of field lines (see Fig. 1).

\item{(iii)} For values of $s \geq 2$, the solutions are physically
acceptable in the limited sense that $v_{\rm min} \leq c$ at all
points (solid and dashed lines in Fig. 2).  For $s < 2$, the solutions
have $v_{\rm min}$ exceeding $c$ beyond a certain value of
$u$. (Technically, we should replace $s$ by $|s|$ in these statements,
but we are interested only in solutions with positive $s$.)

\item{(iv)} Among solutions with $s \geq 2$, the particular solution
with $s=2$ has $v_{\rm min}$ asymptotically tending exactly to $c$
(fast critical surface at infinity), and it is also the only solution
that can be matched to an analytic core on the axis (\S~3.3.3).  In
addition, it is the solution with the minimum torque among all
solutions that have $v_{\rm min} \leq c$.  Thus, this unique solution
is the ``correct'' solution to the self-similar problem.  In a sense,
the constraint on the fast surface and the analyticity argument
provide a physical motivation for Michel's (1969) minimum torque
condition.

\item{(v)} For the solution with $s=2$, the Lorentz factor increases
with distance along the $z$-axis as $\gamma_{\rm min} \sim M
(z/\Rfp)^{1/2}$ (eq. \ref{pargamma}; Beskin \& Nokhrina 2006).
\end{description}

As we show below (and also in the Appendix), the paraboloidal solution
$\nu=1$ acts as a watershed in solution space --- solutions on one
side ($\nu>1$) have very different properties compared to those on the
other side ($\nu<1$).

\begin{figure}
\includegraphics[width=3.3in,clip]{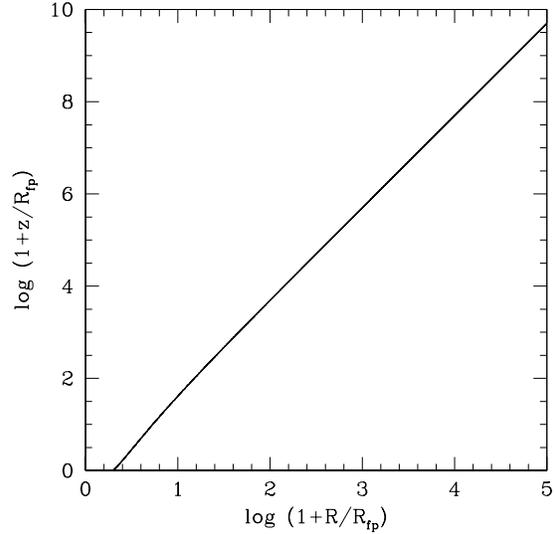}

\caption{Shows the poloidal structure of field lines
for the paraboloidal solution, $\nu=1$.  The calculations correspond
to $M=0.1$ and eight values of $s$: 0.5, 1.0 $\cdots$ 4.0.  All the
solutions have exactly the same poloidal structure (\S~3.1).  Contrast
this with Figs. 3 and 5.}
\end{figure}

\begin{figure}
\includegraphics[width=3.3in,clip]{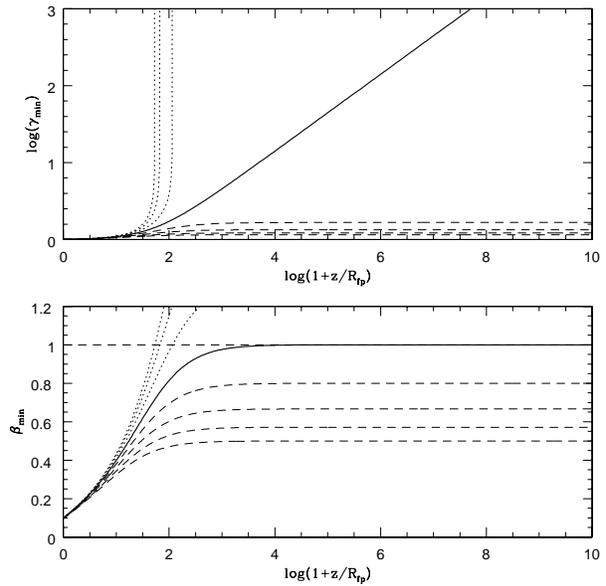}
\caption{Shows the variation of the minimum Lorentz
factor $\gamma_{\rm min}$ and the minimum velocity $\beta_{\rm min} =
v_{\rm min}/c$ as a function of distance $z$ from the disk plane for
the paraboloidal ($\nu=1$) solutions of Fig. 1.  The thick solid line
is the physically interesting solution with $s=2$, whose velocity
asymptotically approaches $c$ at infinity.  The dotted lines
correspond to solutions with $s < 2$, which attain minimum velocities
exceeding $c$.  The dashed lines correspond to solutions with $s > 2$,
whose velocities remain below $c$ at all $z$.}
\end{figure}

\subsection{Solutions with $\nu > 1$}

We have numerically scanned the solution space of equation
(\ref{Tequation}) for the particular choice of parameters $\nu=1.25$
and $M=0.1$.  Keeping these parameters fixed, we solved for $T(u)$
corresponding to a range of values of the eigenvalue $s$.  Then, using
the numerical solution we calculated the components of the magnetic
field, the minimum fluid velocity, the minimum Lorentz factor, etc. We
identified which solutions are physically acceptable, i.e., have
$v_{\rm min} \leq c$ for all $u$ (technically, out to the maximum $u$
to which we integrate, which is $\sim 10^{12}$), and which are not,
which solutions have power-law, i.e., generalized parabolic,
asymptotics (eq. \ref{zvsR}) and which have cylindrical asymptotics as
$u \to \infty$, and we also flagged any special solutions that had
$v_{\rm min} \to c$ as $u \to \infty$.

As per the discussion in \S~2.3, we first checked whether $s^2 \leq
1/(\nu-1)$.  If so, we solved for $u_c$, i.e., the position of the
Alfven surface, by integrating equation (\ref{Tequation}) and applying
the conditions (\ref{bc1}), (\ref{bc2}), (\ref{bc3}).  In practice, we
found it convenient to assume a value of $u_c$ and integrate the
differential equation (\ref{Tequation}) from the Alfven surface
$u=u_c$, where (\ref{bc2}) and (\ref{bc3}) provide initial conditions,
back to $u=0$.  At $u=0$ we checked whether the solution satisfied the
condition $T(0)=1$.  If not, we tried a different value of $u_c$ and
iterated until the solution converged.  We then integrated equation
(\ref{Tequation}) outside the Alfven surface to large values of $u \gg
u_c$ to complete the solution.

If $s^2 > 1/(\nu-1)$, we looked for a solution that lives entirely
inside the Alfven surface.  The only physically acceptable solutions of this kind
are those with cylindrical asymptotics.  Therefore, we used the
discussion in \S~3.5 to assign the necessary boundary conditions at
infinity.  By construction, all of these solutions have $v_{\rm min} <
c$.

Figure 3 shows the poloidal structure of the field lines for various
solutions, and Figure 4 shows the corresponding variations of
$\beta_{\rm min}=v_{\rm min}/c$ and $\gamma_{\rm min}$ with distance.
The thick solid lines correspond to the unique solution that has
$v_{\rm min} \to c$ as $u \to \infty$.  This solution corresponds to
$s \approx 1.6 = 2/\nu$ (see Appendix A.1).  The poloidal shape of
field lines is asymptotically of the form $z \propto R^{2/(2-\nu)}$
(see eq. \ref{zvsR}), i.e., a generalized parabola.

\begin{figure}
\includegraphics[width=3.3in,clip]{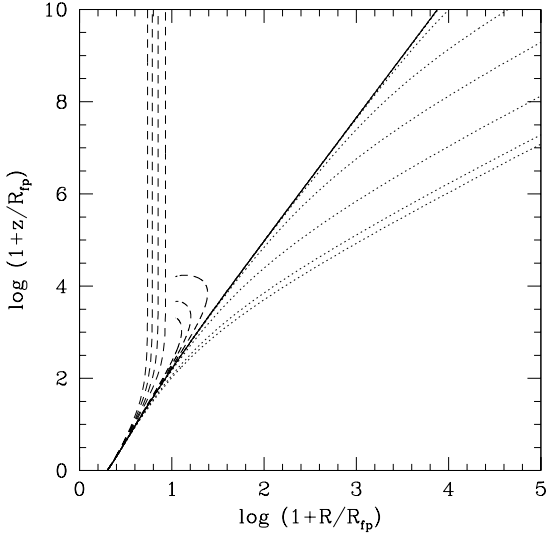}
\caption{Shows the poloidal structure of field lines
for self-similar solutions with $\nu=1.25$.  The calculations
correspond to $M=0.1$ and $s=$ 0.5, 1.0, 1.5, 1.59, 1.599, 1.5999,
1.59996 (dotted lines), $s=1.5999635948$ (heavy solid line), $s$ =
1.7, 1.8, 1.9, 2.5, 3.0, 3.5, 4.0 (dashed lines).  The heavy solid
line is the physically interesting solution that satisfies the
boundary condition (\ref{bc4}).}
\end{figure}

\begin{figure}
\includegraphics[width=3.3in,clip]{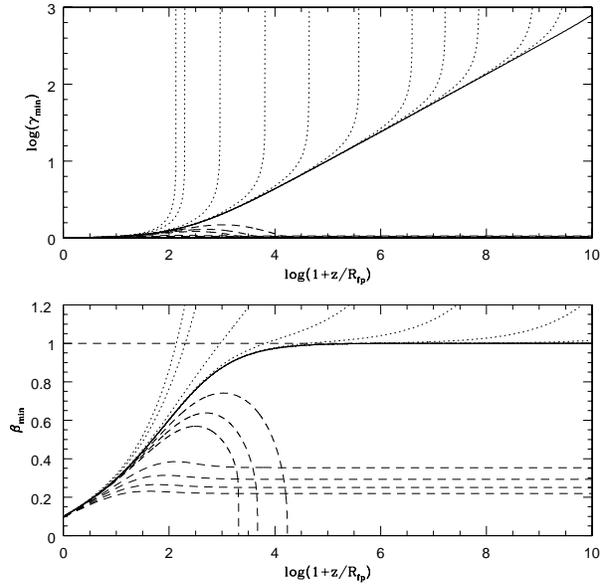}
\caption{Shows the variation of the minimum Lorentz
factor $\gamma_{\rm min}$ and the minimum velocity $\beta_{\rm min} =
v_{\rm min}/c$ as a function of distance $z$ from the disk plane for
the $\nu=1.25$ solutions of Fig. 3.}
\end{figure}

Solutions with $s$ on either side of the above critical value are
``unstable'' (Appendix A.1; see also Contopoulos 1995a).  When $s <
2/\nu$ (shown by dotted lines in Figs. 3 and 4), the solutions have
fluid velocities exceeding $c$ and are unphysical.  They are all
asymptotically conical (see \S~3.5).  Solutions with $s >
1/(\nu-1)^{1/2}$ are entirely inside the Alfven surface and have cylindrical
asymptotics (dashed lines in Figs. 3, 4).  These solutions all have
$v_{\rm min} < c$ as $u\to \infty$.  Over the range $2/\nu < s <
1/(\nu-1)^{1/2}$, field lines go through the Alfven surface, move out to a maximum
radius and then turn back towards the Alfven surface.  The solutions become
singular when they attempt to recross the Alfven surface.

\subsection{Solutions with $\nu < 1$}

\begin{figure}
\includegraphics[width=3.3in,clip]{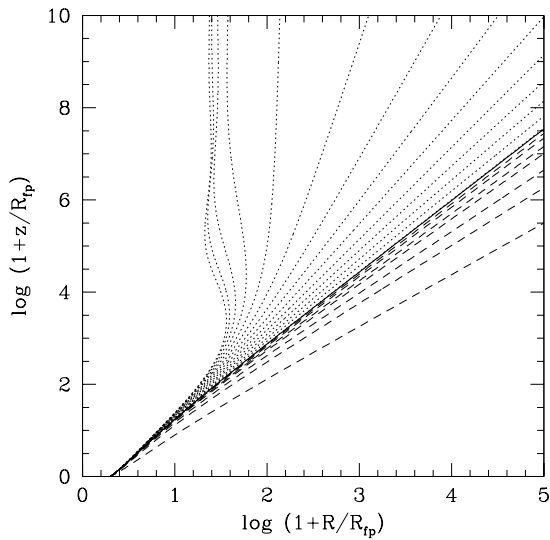}
\caption{Shows the poloidal structure of field lines
for self-similar solutions with $\nu=0.75$.  The calculations
correspond to $M=0.1$ and $s=$ 0.2, 0.4 $\cdots$ 2.8 (dotted lines),
$s=2.8146$ (heavy solid line), $s$ = 2.9, 3.0, 3.2, 3.4, 4.0, 5.0,
10.0 (dashed lines).  The heavy solid line is the physically
interesting solution.  It is the solution with the smallest value of
$s$, i.e., the least amount of toroidal field (minimum torque), that
satisfies the boundary condition (\ref{bc4}).  }
\end{figure}

\begin{figure}
\includegraphics[width=3.3in,clip]{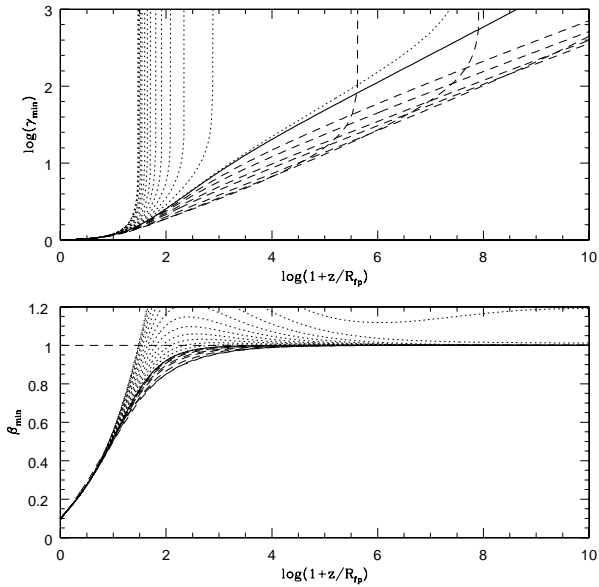}
\caption{Shows the variation of the minimum Lorentz
factor $\gamma_{\rm min}$ and the minimum velocity $\beta_{\rm min} =
v_{\rm min}/c$ as a function of distance $z$ from the disk plane for
the $\nu=0.75$ solutions of Fig. 5.}
\end{figure}

Figures 5, 6 show results corresponding to the parameter choice of
$\nu=0.75$ and $M=0.1$.  As before, we have calculated solutions for a
range of values of $s$ and cataloged their properties.

For $s \leq 1$ (the first five dotted lines in Figs. 5, 6), there are
no asymptotic power-law solutions available (see the discussion in
\S~3.3).  These solutions are asymptotically cylindrical, superposed
with a decaying oscillation, and they are unphysical ($v_{\rm min} >
c$).  For all values of $s > 1$, we have power-law paraboloidal
solutions, and these are ``stable'' in the sense described in Appendix
A.1.

Over the range $1 < s \lesssim 2.8146$, the solutions have $v_{\rm
min} >c$ somewhere within the range of integration over $u$ and are
thus unphysical.  The thick solid lines in Figures 5, 6 correspond to
$s=2.8146$, the lowest value of this parameter for which we can find a
physically allowed solution extending over a large range of $u$ (our
integrations extended out to a maximum $u \sim 10^{12}$).  As per
equation (\ref{zvsR}), the poloidal shape of field lines follows $z
\propto R^{1.55}$. We note that this critical solution has nearly the
same scalings as for the $\nu=1.25$ solution described in the previous
subsection, viz., $s \approx 2/\nu = 2.67$, $z \propto R^{2/(2-\nu)} =
R^{1.6}$, but it is slightly different.  The difference arises because
of the finite value of $M$ that we have selected.  In the limit of
small $M$, it turns out that the critical solution satisfies exactly
the same scalings as the $\nu>1$ solutions described in the previous
subsection (see \S~4.3 below).

For $s > 2.8146$ and up to a second critical value $\sim 3.8$ (the
first four dashed lines in Figs. 5, 6), all the solutions are
physically acceptable, and surprisingly all of them have $v_{\rm min}
\to c$ as $u\to \infty$.  Thus, there is a continuous family of
solutions all of which satisfy the boundary condition (\ref{bc4}).
However, all these solutions have larger toroidal fields, i.e., larger
torques, than the particular solution with $s=2.8146$ described in the
previous paragraph, and their Lorentz factors increase more slowly
with increasing $z$.  Finally, for $s > 3.8$ (the last three dashed
lines), the solutions revert to being unphysical and the fluid
velocity exceeds $c$ over some range of $u$.

We should note that the results of this subsection deviate from those
of Contopoulos (1995a) who states that there are no physically viable
solutions for $\nu<1$ (see his discussion of solutions of type 1).

\subsection{Survey of Solution Space}\label{survey}

The discussion in the previous two subsections was for two specific
values of $\nu$ and for a single choice of $M$.  We now briefly put
these results in a more general context.

\begin{figure}
\includegraphics[width=3.3in,clip]{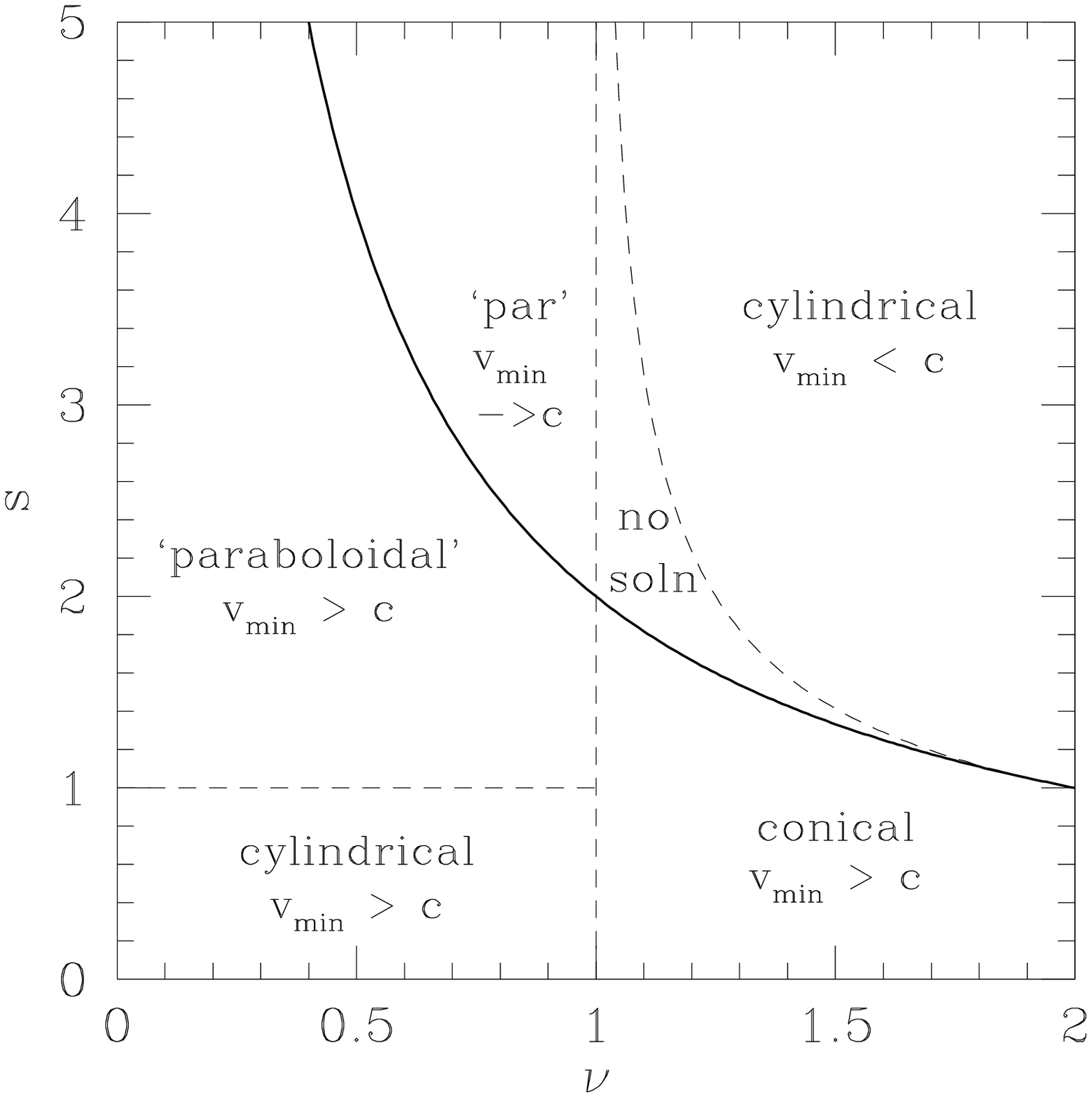}
\caption{Demarcates the different solution regimes
for self-similar force-free winds from a thin disk.  The regimes are
indicated as a function of the similarity index $\nu$ and the field
sweep-back parameter $s$.  The results correspond to the limit $M \ll
1$.  The thick line is the locus of solutions with $s = 2/\nu$.  These
are the physically interesting minimum torque solutions; at each
$\nu$, this is the lowest value of $s$ for which a physically valid
solution with $v_{\rm min} < c$ is possible.  The various regimes
demarcated by dashed lines are described in \S~4.3.}
\end{figure}

We begin by considering the case in which the rotation parameter $M$
is very small.  Figure 7 shows for this case the solution space of
self-similar force-free disk winds as a function of the stream
function index $\nu$ and the field sweep-back parameter $s$ (compare
with Fig. 4 in Contopoulos 1995a, noting that $s\to -H_0 c_n$ and
$\nu\to x$). The thick solid line corresponds to the relation $s =
2/\nu$.

For $\nu > 1$, paraboloidal solutions are available only on the thick
solid line.  Below the line, we have unphysical conical solutions.
Above the line, within a narrow triangular region, we have no
solutions at all.  In this region, solutions cross the Alfven surface
and then try unsuccessfully to re-cross the Alfven surface (\S 4.1).
The solutions are thus unable to reach infinity.  Above this forbidden
zone, there is an extended region of parameter space where there are
asymptotically cylindrical solutions, all of which are physically
consistent.  The main difference between these solutions and those on
the thick solid line is that the cylindrical solutions have $v_{\rm
min} < c$ as $u\rightarrow \infty$.

When $\nu=1$, paraboloidal solutions with physically allowed
velocities are available for all points above and on the thick solid
line.  However, the point $s=2$, which is exactly on the line, is the
only solution that is analytic on the axis (\S~3.3.3), and we have
argued that it is therefore the correct solution.

For $\nu<1$, the situation is quite interesting.  In the limit $M\ll1$
that we are considering here, physically acceptable paraboloidal
solutions are available for all points above and on the thick solid
line.  Moreover, all these solutions have $v_{\rm min}\to c$ for large
$u$ and all are analytic on the axis. However, the solutions with $s =
2/\nu$ (the thick solid line in Fig. 7) are still special in that
these solutions have the minimum torque among all the physical
solutions and also the most rapid acceleration outward.  Previously,
we associated the condition $v_{\rm min}=c$ at infinity with
analyticity on the axis and also with the minimum torque condition.
Now we find a continuum of solutions that have $v_{\rm min}=c$ at
infinity and are analytic, and it is only the requirement of minimum
torque that picks out a unique solution.  Following Michel (1969), we
believe the minimum torque solution is the physically relevant
solution. Points below the thick solid line in Fig. 7 are unphysical
since $v_{\rm min}$ exceeds $c$.  These unphysical solutions are
either paraboloidal or cylindrical, as indicated in Figure 7.

\begin{figure}
\includegraphics[width=3.3in,clip]{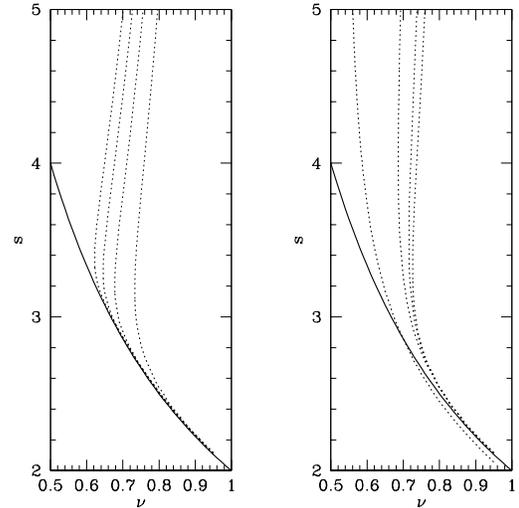}
\caption{Shows further details of self-similar
solutions for $\nu < 1$.  [Left] The thick solid line is the same as
in Fig. 7.  In the limit $M \ll 1$, points on and above this line
correspond to physically consistent solutions and points below are
inconsistent. The four dotted lines show how the region of consistent
solutions shrinks for finite values of $M$; from the left, the lines
correspond to $M=0.0001, ~0.001, ~0.01, ~0.1$.  For each choice of
$M$, the region to the right of the corresponding dotted line is the
region of consistent solutions.  All solutions were numerically
integrated from $u=0$ to $u_{\rm max} \sim 10^{12}$ to determine
whether or not they are consistent.  [Right] The thick solid line is
again the same as in Fig. 7.  The dotted lines show the effect of
reducing the value of $u_{\rm max}$ for $M=0.1$; from the left, the
lines correspond to $u_{\rm max} = 10^3, ~10^5, ~10^7, ~10^9$.  For
each choice of $u_{\rm max}$, the region to the right of the
corresponding dotted line is the region of consistent solutions.}
\end{figure}

When $M$ is not arbitrarily small, the space of physically allowed
solutions is virtually unchanged for $\nu \geq 1$.  However, for
$\nu<1$, the space of allowed solutions shrinks.  This is shown by the
four dotted lines in the left panel of Figure 8, which correspond
(from the left) to $M=0.0001, ~0.001, ~0.01, ~0.1$.  In each case,
only the region to the right of the dotted line is physically allowed,
while the region to the left (even if it is above the thick solid
line) has $v_{\rm min}>c$.  Also, for each value of $\nu$, the lowest
value of $s$ for which a physically allowed solution is present (the
thick segment of the dotted line) is the one with the minimum torque
and the most rapid acceleration.  In effect, this solution takes on
the role of the solution on the thick solid line when $M\ll1$.

We should note that the above statements about physically allowed and
disallowed solutions are based on integrating our solutions out to a
large value of $u \sim 10^{12}$ (beyond this, we are not confident of
the accuracy of the numerical integration.)  However, in practice, one
rarely requires a force-free solution to be physical out to such large
distances since even very high-$\sigma$ relativistic MHD flows are
likely to feel the effects of inertia well before this.  It is
therefore interesting to look for force-free solutions that are
physically reasonable ($v_{\rm min}<c$) out only out to a modest value
of $u$.  The right panel in Figure 8 shows the allowed region in
parameter space for different ranges of integration; from the left,
$u_{\rm max} = 10^3, ~10^5, ~10^7, ~10^9$.  Clearly, solutions are
allowed over more and more of the parameter space as we reduce the
integration range.  However, we still have the situation that, for
each $\nu$ and integration range, the smallest value of $s$ for which
a physically allowed solution is available (the thick segment of the
dotted line), i.e., the minimum torque solution, is the one with the
largest acceleration.

We now come to a key question.  For a given choice of $\nu$ and $M$,
i.e., for given self-similar boundary conditions at the equatorial
plane, do we know which of the many solutions described here is the
{\it correct} solution?  By correct, we mean the solution that would
be picked out by a real system.  We have presented a number of
arguments to suggest that we should pick the solution with $v_{\rm
min}\to c$ since this condition appears to be equivalent to analytic
behaviour on the axis.  But this condition alone is insufficient when
$\nu <1$ since there are many solutions satisfying the condition.  In
the latter case, we have suggested that the correct solution is the
one with the minimum torque.  The numerical simulations described in
\S 5 are designed specifically to verify these assertions.

\subsection{Numerical Verification of Scalings}

In \S\S~3.3.1 and 3.3.2, we wrote down scaling relations for a number
of quantities of interest.  We have verified these scalings by
comparing the analytical results with numerical solutions.

\begin{figure}
\includegraphics[width=3.3in,clip]{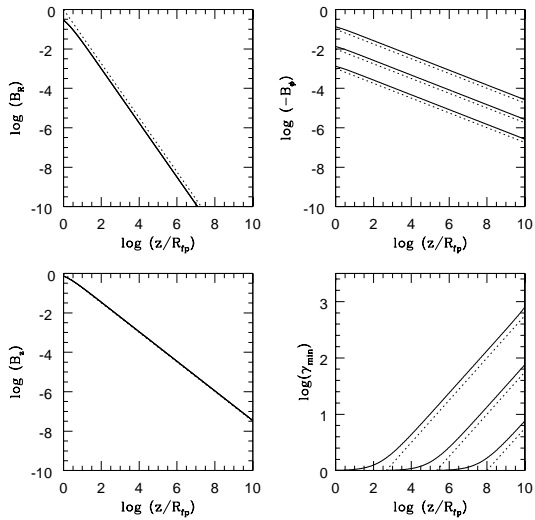}
\caption{Variations of the three field components,
$B_R$, $B_\phi$, $B_z$, and the minimum Lorentz factor $\gamma_{\rm
min}$ as a function of the coordinate $z$ along a field line.  The
index $\nu$ is equal to 1.25, and three solutions are shown,
corresponding to $M=0.1, ~0.01, ~0.001$, with $s = 1.5999635948,
~1.59999998, ~1.60000$, respectively.  The solid lines show the
numerical results and the dotted lines show the approximate scalings
given in \S~3.3.1. }
\end{figure}

Figure 9 shows results for $\nu=1.25$ and three values of $M$: 0.1,
0.01, 0.001.  Shown are the three components of the magnetic field,
$B_R$, $B_\phi$, $B_z$, and the minimum Lorentz factor $\gamma_{\rm
min}$ along a field line.  The solid lines correspond to the numerical
results and the dotted lines show the analytical scalings.  As
discussed in Appendix A.2, the analytical results ignore numerical
coefficients of order unity and so there are small vertical offsets
between the solid and dashed lines.  Apart from this, the agreement is
good.

Note that the poloidal components of the field, $B_R$ and $B_z$, are
essentially independent of $M$.  Thus, rotation has no effect on the
poloidal structure, just as in the paraboloidal case discussed in
\S~3.1.  The toroidal component of the field, $B_\phi$, does depend on
rotation.  In fact, its strength is directly proportional to the
rotation parameter $M$.

The manner in which the Lorentz factor scales with $z$, viz.,
$\gamma_{\rm min} \propto z^{(2-\nu)/2}$, is independent of $M$, but
the actual value of $\gamma_{\rm min}$ at any given $z$ does depend on
$M$.  This is because, with decreasing disk rotation, the acceleration
of the wind starts at a progressively larger value of $z$.  This is
not surprising.  Acceleration is effective only outside the Alfven surface,
because only there does $B_\phi$ become the dominant component of the
field.  When the disk rotates slowly, the Alfven surface is crossed at a larger
value of $z$.

\begin{figure}
\includegraphics[width=3.3in,clip]{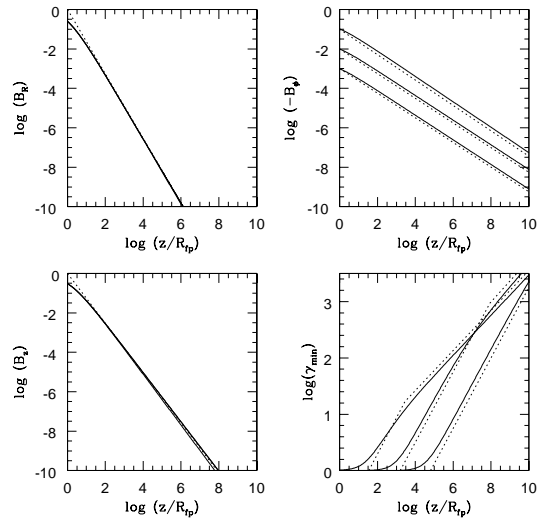}
\caption{Similar to Fig. 8, but for $\nu=0.75$.  The
three solutions correspond to $M = 0.1, ~0.01, ~0.001$, with $s =
2.8146, ~2.67366, ~2.66710$, respectively.  The solid lines show the
numerical results and the dotted lines show the approximate scalings
given in \S3.3.2.}
\end{figure}

Figure 10 shows results for $\nu=0.75$ and the same three values of
$M$.  For each $M$, we have selected the lowest value of $s$ for which
we could obtain a physically valid solution.  As discussed earlier,
this is the solution with the minimum torque and the largest
acceleration.  The poloidal structure of the field shows some
variation with $M$, but the changes are quite small, so we again find
that the poloidal structure is effectively independent of rotation.

The variation of the Lorentz factor with $z$ is rather interesting in
this case.  We see that there are two regimes: a regime of rapid
acceleration soon after a field line crosses the Alfven surface, and a regime of
slower acceleration farther out.  The scalings given in \S~3.3.2 for
$\gamma_{\rm min}$ agree quite well with the numerical results.
Interestingly, the asymptotic Lorentz factor at large $z$ is somewhat
larger for a slower spinning disk than for a rapidly spinning disk.
This is seen both in the numerical results and in the scaling
relations.

\section{Time-Dependent Numerical Simulations}\label{timedep}

In this section, we discuss the results of time-dependent numerical
simulations of force-free nearly self-similar winds.  The general
relativistic force-free electrodynamics code described in
\citet{mckinney2006a} is used to evolve the axisymmetric force-free
equations of motion in a flat space-time in spherical polar
coordinates. This code is an extension of the general relativistic MHD
scheme called HARM \citep{gmt03}, such that the force-free equations
are written as a general set of conservation equations.  In the ideal
force-free degenerate limit, the force-free equations can be written
in a coordinate basis in conservative form as
\begin{equation}
\frac{\partial}{\partial t} (\detg T^t_i) = -\frac{\partial}{\partial
x^j} (\detg T^j_i) + S(\mathbf{T}),
\end{equation}
where the index $i$ runs over all three spatial dimensions and the
index $j$ runs over the two poloidal spatial dimensions and
$\mathbf{T}$ is the stress-energy tensor given by
\begin{equation}
T^{\mu\nu} = b^2 u^\mu u^\nu + \frac{1}{2}b^2 g^{\mu\nu} - b^\mu b^\nu ,
\end{equation}
where $b^\mu$ is the 4-magnetic field in the frame moving at the drift
4-velocity given by $u^\mu$ and the metric $g^{\mu\nu}$ corresponds to
Minkowski space-time with a determinant of $g$.  The quantity
$S(\mathbf{T})$ is a source term that depends on the coordinate
system. These three evolution equations implicitly determine the
evolution of the electric field. The magnetic field is governed by the
induction equations given by
\begin{equation}
\frac{\partial}{\partial t} (\detg B^i) = -\frac{\partial}{\partial x^j} (\detg (b^j u^i - b^i u^j)) ,
\end{equation}
where $B^i$ is the lab-frame magnetic field that obeys the solenoidal
constraint given by
\begin{equation}
\frac{1}{\detg}\left(\frac{\partial}{\partial x^i} (\detg B^i)\right) = 0 .
\end{equation}
For more details see \citet{kom02b,kom04,gmt03,mckinney2006a}.

The force-free version of HARM has been successfully used to study
pulsar magnetospheres \citep{mckinney2006b}, the Blandford-Znajek
split-monopole solution (corresponding to $\nu=0$, McKinney 2006a),
the Blandford-Znajek paraboloidal solution (corresponding to $\nu=1$),
and a Blandford-Znajek-type solution with $\nu=3/4$ \citep{mn06b}. The
MHD version of HARM has been successfully used to study accretion
flows around rotating black holes where nearly force-free jets
self-consistently form and are magnetically accelerated to $\gamma\sim
10$ \citep{gmt03,gsm04,mg04,mckinney2006c,mn06a}.

\subsection{Modelling the Self-Similar Solution}

The divergence of the field strength near the polar axis and at
spherical $r=0$ in the self-similar models is problematic for a
time-dependent numerical code.  Indeed, as we discussed in \S~3.3.3,
we assume that the self-similar solution is modified in a core region
at small cylindrical radii so as to maintain analyticity around the
axis. For analogous reasons, we screen out the divergent region of the
numerical solution near spherical $r=0$ in the simulations by
introducing an artificial ``star.''  The star is centered at $r=0$
with unit radius and is endowed with a constant angular velocity on
its surface equal to the angular velocity in the self-similar disk
model at $R=1$. The poloidal field at the surface of the star is
chosen to be the same as that given by the self-similar solution on
the spherical polar surface $r=1$.

The computational grid is chosen so that at large radii the grid
follows the collimating field lines to ensure good resolution.  The
grid is composed of an arbitrary two-dimensional space with coordinate
directions $\{x_1,x_2\}$.  The $x_1$ grid lines are mapped to the
spherical polar radius $r$ according to
\begin{equation}
r = R_0 + {\rm e}^{x_1^n} ,
\end{equation}
where $R_0=-3$, $n=10$, and the radius of the grid goes from $R_{\rm
in}=1$ to $R_{\rm out}=10^4$ in arbitrary units of length.  The $x_2$
grid lines are mapped to the spherical polar $\theta$ according to
\begin{equation}
\theta =
\left(\frac{\pi}{2}\right)\left(\frac{1+\tan^{-1}[h(x_2-1/2)]}{\tan^{-1}[h/2]}\right)
,
\end{equation}
where
\begin{equation}
h(r) = \left(\frac{r-r_0}{r_1}\right)^\alpha ,
\end{equation}
and we choose $r_0=0$, $r_1 = 10$.  The index $\alpha$ is chosen to be
the same as the index in the scaling of the opening angle of a field
line at large radii in the self-similar solution, $\theta_j\propto
r^{-\alpha}$.  From the analytic asymptotic solution given in equation
(\ref{muLC+}), we find that $\alpha = 1/s$.

The resolution for all models is chosen to be $256\times 128$.  The
solutions are well-converged compared to low resolution models except
very close to the poles where the coordinate singularity of spherical
polar coordinates causes minor artifacts that do not affect the
results.

\subsection{Obtaining a Force-Free Stationary Solution}

Steady state force-free numerical solutions with no discontinuities or
surface currents above the disk surface are found by choosing boundary
conditions determined by an analysis of the Grad-Shafranov equation
(see, e.g., \citealt{bogo97,beskin97}).  For solutions that pass
through the Alfven surface at some radius, one is required to fix the
magnetic field component perpendicular to the conductor ($B_r$ for the
star and $B_\theta$ for the disk) and to specify two other constraints
at the star or disk, viz., $E_\phi=0$, and the value of $\Omega_F$,
the field line angular velocity.

For axisymmetric, stationary solutions, the frozen-in condition of
ideal MHD implies that the field line velocity $v_i$ is completely
determined by the field $B_i$ and field rotation frequency $\Omega_F$
(see equation 46 in \citealt{mckinney2006a}).  Thus, during the
simulation the 3-velocity at the stellar surface and on the disk
equatorial plane is set to agree with this condition.  Such a
3-velocity is generally time-like for points inside the Alfven
surface, but outside this region the 3-velocity can sometimes be
space-like and unphysical (this is analogous to $v_{\rm min} > c$
which is discussed extensively in previous sections).  In the event
that the 3-velocity becomes space-like during the simulation, the
Lorentz factor is locally constrained to a fixed large value as
described in section 2.5 of \citet{mckinney2006a}. This safety feature
is necessary to handle the violent evolution seen early in the
simulations, but it is usually not activated once the solution
approaches a stationary state.

In the work described here, the initial conditions in the disk are
chosen to correspond to the chosen self-similar solution, described by
the parameters $\nu$, $M$ and $s$.  The only difference is that we
initially set $B_\phi=0$ and allow the code to develop whatever
toroidal field it wishes. During the time evolution, $B_\theta$ at the
disk is held fixed at its initial self-similar value, and the field
angular velocity $\Omega_F$ is set according to the self-similar
model.  On the star, the self-similar value of $B_r$ is held fixed
within $r=1$ and $\Omega_F(r=1)$ is set equal to $\Omega_F(R=1,z=0)$.
Thus the stellar and disk values of $\Omega_F$ match at $R=1, ~z=0$.

The initial non-rotating state is far from steady state, so the model
undergoes violent non-stationary evolution. Eventually, however, it
relaxes to a steady state.  All solutions thus found are necessarily
stable to Eulerian axisymmetric perturbations.

\subsection{Models with $\nu=1$}

The self-similar model with $\nu=1$ corresponds to the paraboloidal
model of \citet{blandford76}.  As we have seen, the poloidal structure
of the field is independent of $M$ and $s$.  We initialize the
simulation with the analytical poloidal solution and set the angular
velocity profile in the disk to correspond to the desired value of
$M$, viz., $M=0.25$. From the discussion in the previous sections, we
know that physically allowed solutions are available for all $s \geq
2$, of which the solution with $s=2$ is the one that (i) has $v_{\rm
min} =c$ at large distance, (ii) is analytic on the axis and (iii) has
minimum torque. However, recall that $B_\phi$ is set equal to 0
initially (corresponding to $s=0$) and we allow the code to evolve to
whatever value of $s$ it chooses. Which value of $s$ does the
time-dependent code select?

\begin{figure}
\includegraphics[width=3.3in,clip]{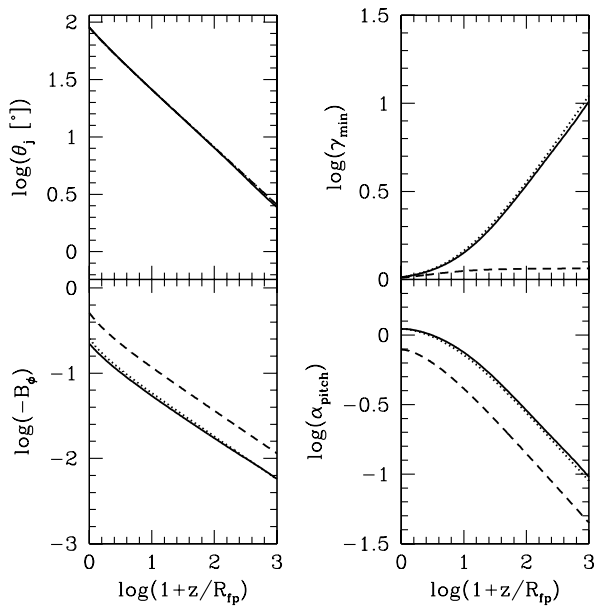}
\caption{Shows for $\nu=1$, $M=0.25$, the run of various quantities as
a function of $z$ following a particular field line with foot-point at
$R_{\rm fp} = 2.0$.  The four panels correspond to (i) the opening
angle of the field line ($\theta_j=\tan^{-1} R/z$), (ii) the minimum
Lorentz factor ($\gamma_{\rm min}$), (iii) the toroidal field strength
($B_\phi$), and (iv) the pitch angle ($\alpha_{\rm pitch}= \tan^{-1}
B_p/|B_\phi|$).  In each panel, the solid line corresponds to the
final converged solution obtained from the time-dependent numerical
simulation, the dotted line corresponds to the analytical self-similar
solution with $s=2$, and the dashed line corresponds to the analytical
solution with $s=4$.  It is clear that the numerical solution agrees
very well with the $s=2$ analytical solution.  This is the smallest
value of $s$ for which a physically consistent ($v_{\rm min}\le c$)
solution is available, and it is the minimum torque solution.}
\label{jetstructnu1}
\end{figure}

Figure~\ref{jetstructnu1} shows the results of the numerical model
compared against the self-similar analytical model.  The plot shows
the dependence of various quantities along the particular field line
that starts at a foot point radius of $R_{\rm fp}=2.0$.  The top left
panel shows the field line's angle away from the polar axis as a
function of distance from the disk plane.  This quantity is purely a
function of the poloidal structure of the field lines, and it agrees
very well with the self-similar solution.  The other panels show the
minimum Lorentz factor, the toroidal field strength, normalized such
that $B_z(R=1,z=0)=1$, and the pitch angle of the field line defined
as
\begin{equation}\label{pitchdef}
\alpha_{\rm pitch} \equiv \tan^{-1}\left(\frac{B_p}{|B_\phi|}\right) ,
\end{equation}
where $B_p$ is the poloidal field strength.  For comparison, the
dotted lines show the analytical results for the case $s=2$ and the
dashed lines show the corresponding results for $s=4$ (as a
counter-example).  It is clear that the numerical simulation converges
to the model with $s=2$, for which we have
\begin{equation}\label{goodpitch}
\alpha_{\rm pitch} \approx \tan^{-1}\left(\frac{c}{R\Omega_F}\right)
\end{equation}
at large radii.

\begin{figure}
\includegraphics[width=3.3in,clip]{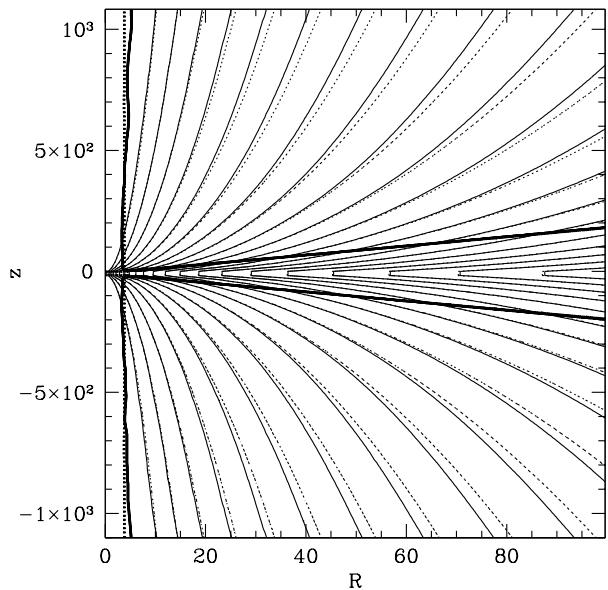}
\caption{Contours of the stream function $P$ showing the poloidal
structure of field lines for $\nu=1$, $M=0.25$.  The thin solid lines
show the numerical solution and the dotted lines show the analytical
self-similar solution for $s=2$.  Note the good agreement.  The thick
lines are explained in the caption to Fig.~\ref{compfieldnu1zoom}.}
\label{compfieldnu1}
\end{figure}

\begin{figure}
\includegraphics[width=3.3in,clip]{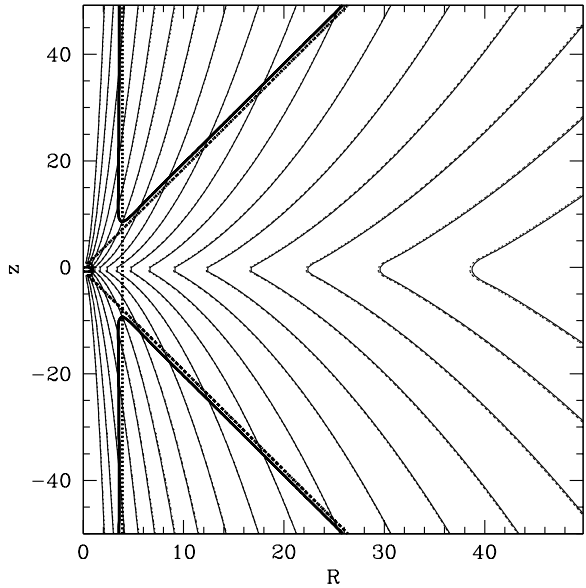}
\caption{Close-up of the central regions of Fig.~\ref{compfieldnu1}
highlighting the location of the Alfven surface, shown by the thick
solid line.  Ignoring the disk, the uniformly rotating star would have
its Alfven surface at $R=4$, shown by the thick vertical dotted line.
The pure analytical self-similar disk wind solution has its Alfven
surface at a fixed angle of $\theta=0.49$ radians away from each polar
axis, as shown by the thick sloping dotted lines.  It is seen that the
Alfven surface in the numerical solution smoothly interpolates between
the two analytical Alfven surfaces.} \label{compfieldnu1zoom}
\end{figure}

Figure~\ref{compfieldnu1} shows the simulation results for the
poloidal field geometry and the location of the Alfven surface.
Overlapping the simulation results is the analytical self-similar
field geometry and the location of the self-similar Alfven surface.
There is excellent agreement for the Alfven surface and reasonable
agreement for the disk field lines. Because the simulation has a
rigidly rotating star at the center, there is a second branch of the
Alfven surface corresponding to the stellar rotation.  The position of
this branch is in reasonable agreement with the analytical estimate,
with minor variations due to under-resolution of the Alfven surface at
large radius.

Figure~\ref{compfieldnu1zoom} shows the inner region of
Figure~\ref{compfieldnu1} in order to show how the components of the
Alfven surface associated with the star and the disk merge into a
single Alfven surface.  This figure also shows that the field lines
are in excellent agreement at small radii.

\subsection{Models with $\nu=0.75$}

Unlike the case of $\nu=1$, self-similar models with $\nu=0.75$ have
different poloidal field geometries for each value of $M$ and $s$. The
values of $B_\theta$ and $\Omega_F$ in the disk are specified by the
boundary conditions, i.e., the parameters $\nu$ and $M$, and are of
course independent of $s$.  For this study we choose $M=0.1$ and
initialize the simulation with the poloidal solution corresponding to
two values of $s$: $2.8146$, corresponding to the minimum torque
condition, and $5.0$, a much larger value.  Which value of $s$ will
the time-dependent numerical simulation choose?

\begin{figure}
\includegraphics[width=3.3in,clip]{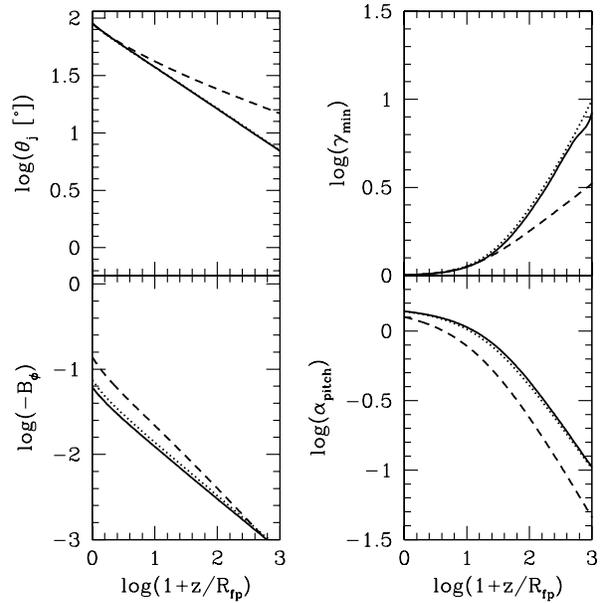}
\caption{Similar to Fig.~\ref{jetstructnu1}, but for $\nu=0.75$.  The
numerical model (solid lines) agrees well with the analytical
self-similar solution with $s = 2.7$ (dotted lines), but not with the
self-similar solution with $s=5.0$ (dashed lines). The former value of
$s$ is approximately the smallest for which a physically consistent
solution ($v_{\rm min}\le c$) is possible, i.e., it is the minimum
torque solution.} \label{jetstructnu3o4}
\end{figure}

Figure~\ref{jetstructnu3o4} shows the same information as
Figure~\ref{jetstructnu1}, but for $\nu=0.75$.  Regardless of which
solution we use to initialize the simulation, the final state selected
by the time-dependent code corresponds to a solution with $s\approx
2.7$.  This is close enough to the special value of $s=2.8146$ that we
claim the final solution is the minimum torque solution, in agreement
with the proposal of Michel (1969).  As discussed in
section~\ref{survey}, for finite $M$ and for $s$ below a critical
value, there is a finite radius beyond which the solution becomes
unphysical. Indeed, $s\approx 2.7$ leads to $v_{\rm min}=c$ at a
finite radius somewhat smaller than the size of the simulation box.
The simulation does reach $v_{\rm min}=c$ for $r\sim 10^3$, although
the numerical method would have to be improved to verify that this is
not just numerical error.

\subsection{Models with $\nu=1.25$}

Like self-similar models with $\nu=0.75$, models with $\nu=1.25$ have
different poloidal field geometries for each value of $M$ and $s$,
while the values of $B_\theta$ and $\Omega_F$ in the disk are the same
for each $s$.  For this study we choose $M=0.4$ and initialize the
simulation with the self-similar poloidal solution corresponding to
two values of $s$: $1.6$, which corresponds to the paraboloidal
solution on the thick solid line in Figure 7, and $s=2.5$, which
corresponds to an asymptotically cylindrical solution.  Again, the
interesting question is which solution will be picked by the
time-dependent numerical simulation.

\begin{figure}
\includegraphics[width=3.3in,clip]{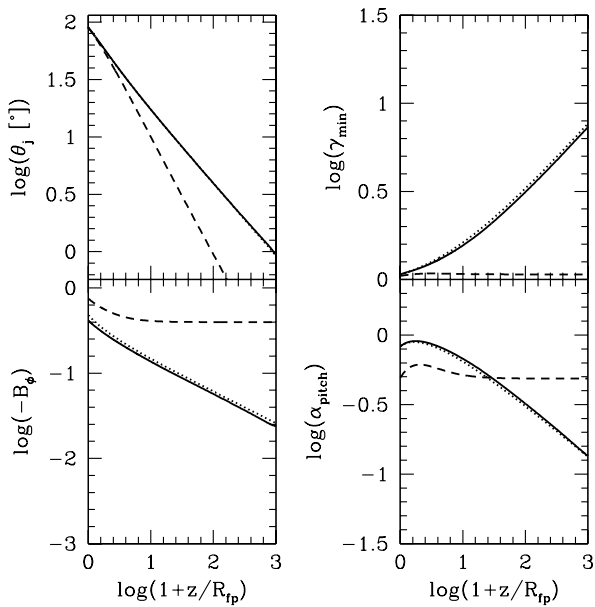}
\caption{Similar to Fig.~\ref{jetstructnu1}, but for $\nu=1.25$.  The
numerical model (solid lines) agrees well with the analytical
self-similar solution with $s = 1.6$ (dotted lines), but not with the
self-similar solution with $s=2.5$ (dashed lines). The former value of
$s$ is the smallest for which a physically consistent solution
($v_{\rm min}\le c$) is possible, i.e., it is the minimum torque
solution.} \label{jetstructnu5o4}
\end{figure}

Figure~\ref{jetstructnu5o4} shows the same information as
Figure~\ref{jetstructnu3o4}, but for $\nu=1.25$.  Clearly the
time-dependent code has chosen a solution close to the paraboloidal
solution with $s\approx 1.6$.  Even when we start the simulation with
the cylindrical solution, the field lines spontaneously open out and
become paraboloidal with $s=1.6$.  The final solution is the minimum
torque solution for this problem.

Note that the behaviour at large radius of the time-dependent solution
that starts with $s=2.5$ is difficult to follow because the solution
undergoes quite a large change from cylindrical to paraboloidal
streamlines. Thus one under-resolves either the initial solution or
the evolved final solution at large radii. For such models we
performed two simulations with a grid that resolves one or the other
type of field geometry, and found that both resulted in the same final
solution with $s\approx 1.6$.

\subsection{Summary of Time-Dependent Simulation Results}

In summary, the time-dependent force-free numerical simulations reveal
the following:

(i) For a given choice of $\nu$ and $M$, a self-similar force-free
system spontaneously seeks the smallest value of the eigenvalue $s$
for which a physically consistent solution ($v_{\rm min} \leq c$) is
possible throughout the finite numerical grid.  Since $s$ measures the
strength of the toroidal field, an equivalent statement is that the
system chooses the configuration that has the least sweepback of field
lines in the toroidal direction, or the minimum torque.

(ii) The final converged numerical solution matches very well the
analytical self-similar solution that corresponds to this value of
$s$.  There is close agreement in the profiles of the components of
the magnetic field, $B_R$, $B_\phi$, $B_z$, and the minimum Lorentz
factor $\gamma_{\rm min}$.

\section{Summary and Discussion}

With a view ultimately to understanding the nature of MHD outflows
from relativistic accretion disks around spinning black holes, we have
considered in this paper the much simpler problem of a self-similar
force-free outflow from an infinitely thin rotating equatorial disk.
The problem is mathematically described in terms of two dimensionless
numbers: (i) $\nu$, the radial index of the stream function $P$ on the
equatorial plane (eq. \ref{Peq}), which is defined such that the
magnetic field scales as $R^{\nu-2}$ on this plane; (ii) $M$, a
rotation parameter, defined such that the azimuthal rotation velocity
on the equatorial plane is $Mc$, independent of $R$ (eq. \ref{Omeq}).
Given these two parameters, the self-similar problem naturally leads
to two dimensionless eigenvalues: (i) $s$, which describes the degree
of sweepback of the magnetic field lines at the equatorial plane (eq.
\ref{sdef}); (ii) $u_c$, which gives the value of $z/R$ at the Alfven
surface (also called the light cylinder in the pulsar magnetosphere
problem).  The eigenvalues are obtained by solving the differential
equation (\ref{Tequation}) with appropriate boundary conditions.

Although this is an extremely simple model, the solution space is
nevertheless surprisingly rich.  We find generalized paraboloidal
solutions (\S~3.3), cylindrical solutions (\S~3.5), conical solutions
(\S~3.6), and even converging solutions (\S~3.4).  Moreover, some of
the solutions have field lines crossing the Alfven surface, while others live
entirely inside the Alfven surface.  Numerical examples of the different kinds of
solutions are shown in Figures 1--6, and the regions of parameter
space over which the different kinds are found are indicated in
Figures 7 and 8.  In \S\S~3.3.1, 3.3.2 we discuss scaling relations
for paraboloidal solutions, and in Figures 9 and 10 we show that these
analytical scalings agree well with numerical solutions.

Given the bewildering variety of solutions, we are faced with the
question: which is the ``correct'' solution for a given choice of
$\nu$ and $M$?  In an attempt to make some headway on this question,
we may wish to impose the condition that the minimum fluid velocity
$v_{\rm min}$, as defined in \S~2.4, should not exceed $c$ anywhere
within the flow.  However, this condition only eliminates half the
parameter space of solutions (see Figs. 7, 8). Requiring in addition
that the solution should be analytic on the axis, we find this
corresponds to the desirable condition $v_{\rm min} \to c$
asymptotically far from the disk (\S~3.3.3).  This leads us to the
stronger condition $v_{\rm min} = c$ at infinity instead of just $\leq
c$, and it gives us a unique solution for all $\nu \geq 1$. However,
for $\nu < 1$, it still leaves us with a continuous family of
physically valid solutions which spans, for each choice of $\nu$ and
$M$, a range of values of $s$.  To break this residual degeneracy, we
are compelled to invoke yet another condition, viz., that the correct
solution is the one with the minimum toroidal field, i.e., the minimum
torque (Michel 1969).  This latter condition subsumes the $v_{\rm
min}\to c$ condition, and gives a unique solution for all values of
$\nu$.  It is, however, less well motivated than the analyticity
condition on the axis or the requirement that the fast surface be
located at infinity.  We are thus led to use axisymmetric numerical
simulations to verify which solution is picked out by the
time-dependent system.

\S~5 describes these numerical simulations and summarizes the results
(see Figs. 11--15).  For each of the three representative combinations
of \{$\nu, ~M$\} that we analyze, we find that the numerical models
converge to a unique final steady state, which is independent of the
initial starting state.  In all cases, the final state has a value of
$s$ equal to the smallest value of this parameter for which our
analytical model gives a physically consistent solution (with $v_{\rm
min}\le c$ over the domain of interest), i.e., the solution with the
minimum torque.  This is also the solution with the largest
acceleration.

One interesting question is prompted by the numerical simulations.  In
the mathematical analysis described in this paper and also in the
physical discussion of the previous paragraphs, we assigned great
importance to the point at infinity, or at the edge of the grid, and
assumed that one or more important boundary condition is set there.
This is a natural consequence of working with the steady state problem
which involves analyzing the elliptic Grad-Shafranov equation.  The
numerical simulations, however, work with the hyperbolic
time-dependent force-free equations.  Surprisingly, in the numerical
simulations, the flow chooses the more-or-less correct configuration
for the magnetic field near the equatorial plane long before any
signals (moving at the speed of light) have reached the outer boundary
of the grid, or even reached the Alfven surface.  Thus, the
time-dependent problem appears not to require the point at infinity to
set up the local flow. Exactly how does it do this?  One answer is
that the condition $v_{\rm min} \to c$ at infinity is equivalent to
analyticity on the axis, i.e., the absence of a singular current on
the axis (\S~3.3.3).  The numerical code forbids a singular current on
the axis (at all $z$) and would therefore naturally pick out such
solutions. For $\nu<1$, we require an additional condition that the
solution has the minimum torque (or maximum acceleration). This again
is a condition that could, in principle, be applied at the equatorial
plane and perhaps this is how the code is able to pick the correct
solution promptly.

Ultimately, we suspect that only a perturbation analysis of the
self-similar steady solutions we have obtained in this paper will
provide real understanding of which solution is correct for each
situation. It could be that the minimum-torque/maximum-acceleration
solution is the only one that is stable to axisymmetric perturbations.

The quasi-analytic work we presented is closely related to that of
Contopoulos (1995a), who investigated self-similar force-free flows in
order to better understand MHD solutions obtained by Li et al. (1992)
and Contopoulos (1994).  Our work differs from Contopoulos (1995a) in
at least two ways: (i) We have identified a larger set of solutions ;
(ii) Contopoulos (1995a) suggests that the value of $-sM$ ($H_0$ in
his notation) has to be fine-tuned in order to obtain solutions that
reach asymptotic infinity with high Lorentz factors. For a physical,
time-dependent solution, he suggests that this fine-tuning corresponds
to the requirement that waves propagate from asymptotic infinity and
reach the origin, and this would take an infinite amount of time.  He
suggests, therefore, that such solutions are inapplicable as physical
models.

On the other hand, following Blandford (1976) for $\nu=1$, we have
been able to identify a regularity condition on the polar axis that
leads to a unique solution for each $\nu\ge 1$, and we use the
minimum-torque (minimum-energy) condition to choose a unique solution
for each $\nu<1$.  As discussed above, we confirmed these conditions
as natural by performing time-dependent numerical simulations that
generate solutions that are, by construction, regular on the polar
axis and necessarily stable. These simulations indicate that the
minimum-torque condition is the natural condition and that there is no
fine-tuning required to obtain solutions that reach asymptotic
infinity with $v_{\rm min}\to c$. However, for $M$ close to unity the
solutions with $\nu<1$ are indeed sensitive to the value of $s$ (Fig.
8). As applied to astrophysical winds, this may indicate that for
disks with $\nu<1$ the force-free approximation is difficult to
maintain at arbitrarily large distances.

Another important issue is how the self-similar force-free solutions
described in this paper are related to self-similar MHD solutions such
as those discussed by Li et al. (1992) and Vlahakis \& Konigl (2003).
We anticipate that MHD outflows that start out with high values of the
magnetization parameter $\sigma$ will closely follow one of our
force-free solutions out to a certain distance before plasma inertia
induces deviations.  Exactly which of our force-free solutions is
picked by the MHD flow?  Following the work of Goldreich \& Julian
(1970), one suspects that MHD would automatically single out the
minimum torque force-free solution, but this remains to be seen.  It
is also interesting to ask exactly where the MHD solution would start
deviating from the force-free solution.

Returning to the results obtained in this paper, we note that, to a
good approximation, the thick solid line in Figure 7 gives the minimum
torque solution for any $\nu$ and $M$.  The eigenvalue $s$ is given
approximately by $s \approx 2/\nu$, and the magnetic field components
and the Lorentz factor have the approximate asymptotic scalings given
in \S\S~3.3.1, 3.3.2.  Some of these results have been derived by
other authors (e.g., Beskin \& Nokhrina 2006), but some appear to be
new.  The unification of $\nu \geq 1$ models and $\nu < 1$ models is
another relevant contribution of the present paper.  Our recent work
(McKinney \& Narayan 1995a,b) has shown that GRMHD models have
currents and fields described by $\nu\approx 3/4$ for a wide range of
conditions.  It is therefore important to understand the properties of
self-similar models with this particular value of the index,
originally introduced by Blandford \& Payne (1982).

An interesting result of our analysis is that rotation has almost no
effect on the poloidal structure of field lines.  This was known to be
the case for the split monopole solution of Michel (1973a) and the
$\nu=1$ solution of Blandford (1976), but we now find that it is true
for the entire family of self-similar solutions considered in this
paper.  It suggests that the collimation of astrophysical jets is not
the result of the toroidal field associated with rotation.  In our
solutions, collimation seems to be produced by the poloidal field
itself.  In effect, each field line is collimated by the pressure
associated with field lines further out.  However, this result is for
the specific self-similar model we have considered, which has a flat
rotation curve.  It remains to be seen if the results carry over to a
disk with a Keplerian rotation profile.

In contrast to the case of collimation, our models show that rotation
and toroidal field are critical for {\it accelerating} the force-free
wind.  Serious acceleration begins only when the toroidal component of
the magnetic field dominates over the poloidal component, which
happens only after a field line crosses the Alfven surface.  The
larger the rotation of the disk, the closer the Alfven surface is to
the foot-point of a field line, and the sooner strong acceleration is
initiated. This is illustrated in Figures 9 and 10.

Finally, we note that the force-free winds we have considered in this
paper are highly idealized, and their relevance to real disk winds is
unclear since MHD turbulence within the disk may tangle up such
large-scale fields \citep{mckinney2005,mn06b}.  Our hope is that some
of the analytical results and qualitative insights obtained here may
carry over to more realistic MHD models of winds.

\section*{Acknowledgments}

We are grateful to Vasily Beskin and the referee, Serguei Komissarov,
for a detailed reading of the paper and for numerous helpful comments.
This work was supported in part by NASA grant NNG04GL38G. JCM was
supported by an Institute for Theory and Computation Fellowship and
AJF by a Harvard Junior Fellowship.

\appendix
\section{Analysis of Paraboloidal Solutions}

In this Appendix we focus on paraboloidal solutions and derive a few
asymptotic results in the limit of large $u = z/R$.  We begin by
separating out equation (\ref{Tequation}) into terms of different
orders:
\begin{eqnarray}
u^2 TT'' &-& u^2{T'}^2/\nu +(3-2\nu)uTT' +\nu(\nu-1)(1-s^2) T^2 \nonumber\\
&-& M^{-2} T^{(\nu+2)/\nu} \left[u^2 T'' + (3-2\nu)u T' + \nu(\nu-2)
T\right]
\nonumber\\
&+& TT'' - {T'}^2/\nu \nonumber\\
&-&  M^{-2} T^{(\nu+2)/\nu} T'' = 0. \label{Teq2}
\end{eqnarray}
By ``paraboloidal'' solutions, we mean those in which $T(u)$ goes as
$u^{-\mu}$ with a positive value of $\mu$ as $u \to \infty$.  We will
concentrate on the region well outside the Alfven surface, where $T(u)
\ll M^\nu$.  It is obvious that the dominant terms in equation
(\ref{Teq2}) in this region are those listed in the first line, and so
we begin with an analysis of these terms.  The single term in the
fourth line is always much smaller than the others and we will neglect
it.  The terms in lines 2 and 3 are of intermediate importance and we
consider them as needed.

The first line of equation (\ref{Teq2}) is homogeneous and is
identical to equation (\ref{asympeq}) in the main text.  It has two
independent power-law solutions.  We require a decaying solution, and
such a solution exists only when $s > 1$ (as in the main text, we
restrict our attention to positive values of $M$ and $s$).  The
solution is given by
\begin{equation}
T(u) \sim u^{-\mu}, \qquad \mu = \nu (s -1). \label{mueq}
\end{equation}
In the following two subsections, we consider the effect of small
perturbations on this solution and calculate the next order term in a
power series expansion.

\subsection{Perturbation Analysis of the Homogeneous Equation}

Focusing still on just the homogeneous equation represented by the
first line of (\ref{Teq2}), consider small perturbations around the
power-law solution (\ref{mueq}):
\begin{equation}
T(u) \approx u^{-\mu} + \epsilon(u), \qquad \mu = \nu (s-1), \qquad u
\gg u_c. \label{epsilon}
\end{equation}
We will assume that $\epsilon(u) \ll u^{-\mu}$.  Substituting
(\ref{epsilon}) in (\ref{Teq2}) and retaining only the leading terms,
we obtain
\begin{eqnarray}
0&=&u^2\epsilon'' + \left({2\mu\over\nu}+3-2\nu\right) u
\epsilon' \nonumber\\
&+&\left[\mu(\mu+1)-(3-2\nu)\mu +2\nu(\nu-1)(1-s^2)\right] \epsilon .
\label{epseq}
\end{eqnarray}
This equation has power-law solutions of the form $\epsilon \sim
u^{-\xi}$, with two values of $\xi$:
\begin{equation}
\xi_1 = \mu, \qquad \xi_2 = \mu \left({2\over\nu}-1\right) +2(1-\nu).
\end{equation}
The first solution is trivial --- it merely reproduces the leading
order term in equation (\ref{epsilon}) and carries no new information.
The second solution is, however, interesting.  It gives an asymptotic
behaviour for $T(u)$ of the form $u^{-\mu} (1 + C u^{-(\xi_2-\mu)})$,
where $C$ is a constant and
\begin{equation}
\xi_2 - \mu = 2 (1-\nu) \left({\mu\over\nu}+1\right).
\end{equation}

We now consider the stability of the solution; we mean stability in a
mathematical sense, not dynamical.  The perturbation $\epsilon(u)$
will remain smaller than the primary term $u^{-\mu}$ with increasing
$u$ only if $\xi_2 - \mu > 0$, i.e., only if $\nu < 1$.  Thus, so long
as $\nu<1$, the paraboloidal solution is ``stable'' for any positive
value of $\mu$, i.e., for any $s > 1$.  This is confirmed by Figure 5,
where we see that all the solutions with $s > 1$ are well-behaved and
have power-law, i.e., paraboloidal, behaviour as $u \to \infty$.

When $\nu > 1$, however, we see that $\xi_2-\mu$ is always negative,
and so perturbations grow with increasing $u$ and ultimately dominate
over the leading order term $u^{-\mu}$.  This is a sign that the
solution is ``unstable,'' as confirmed by Figure 3.  Apart from the
cylindrical solutions, we see that nearly all the other solutions are
either conical, i.e., $T(u)$ goes to 0 at a finite $u=u_f$, or
singular, i.e., $T(u)$ attempts, and fails, to recross the Alfven
surface. Obviously, none of these solutions is physically
satisfactory. There is, however, one (and only one) paraboloidal
solution which manages to avoid the instability; it is shown as the
thick solid line in Figure 3.  To understand how this special solution
manages to avoid growing perturbations, we note that perturbations to
the primary solution $u^{-\mu}$ are generated by the terms in lines
2--4 of equation (\ref{Teq2}).  So let us briefly take a look at these
terms.

Consider the terms in the second line of equation (\ref{Teq2}).  They
add up to zero when $T(u)$ has the power-law form $u^{-(2-\nu)}$.
Thus, for this particular power-law, these terms do not introduce any
perturbations at all to the solution.  In general, the primary
power-law solution $u^{-\mu}$ is not of this special form.  However,
there is one particular case when the same power-law solution
satisfies both lines 1 and 2 of equation (\ref{Teq2}).  This is when
\begin{equation}
s = {2\over\nu}, \qquad \mu = 2-\nu. \label{special}
\end{equation}
So for this one value of $s$, line 2 does not introduce any
perturbations to the primary $u^{-\mu}$ dependence of $T(u)$.  We
would therefore expect the solution for this particular value of $s$
to be stable.  It is gratifying to see that the special solution shown
by the thick solid line in Figure 3 does indeed correspond to $s$
being very nearly equal to this special value, and the solution does
behave asymptotically as $u^{-(2-\nu)}$.  Of course, we only
considered the first two lines of (\ref{Teq2}).  For the above special
solution, line 3 is of a lower order than line 2, but it is not
completely negligible.  Therefore, the paraboloidal solution does not
correspond exactly to the value of $s$ given in (\ref{special}), but
it is very close.

In summary, for $\nu < 1$, a paraboloidal solution is available for
all $s > 1$, while for $\nu > 1$ it is available for only one specific
value of $s$ which is nearly equal to $2/\nu$.

\subsection{Power Series Expansion and Lorentz Factor}

In this subsection, we ignore line 2 of equation (\ref{Teq2}).  When
$\nu>1$, we showed above that the contribution of this line vanishes
for the only paraboloidal solution available.  When $\nu < 1$, this
line is sub-dominant relative to line 3 for most paraboloidal
solutions of interest to us, especially the ones on and above the
thick solid line in Figure 7.

Consider, therefore, the differential equation consisting of lines 1
and 3 of equation (\ref{Teq2}).  As before, consider a solution of the
form (\ref{epsilon}), where we assume that the second term is much
smaller than the first.  When this solution is substituted in line 1,
the zeroth order terms cancel and the first order terms take the form
given in the left hand side of equation (\ref{epseq}).  On the other
hand, the zeroth-order terms give a non-vanishing contribution when
substituted in line 3.  Thus, collecting terms, we obtain the
following differential equation for $\epsilon$:
\begin{eqnarray}
&~&u^2\epsilon'' + \left({2\mu\over\nu}+3-2\nu\right) u
\epsilon'\nonumber\\
&+&\left[\mu(\mu+1)-(3-2\nu)\mu + 2\nu(\nu-1)(1-s^2)\right] \epsilon
\nonumber \\ &=& -\left[\mu(\mu+1) - {\mu^2\over\nu}\right]
u^{-\mu-2}, \qquad \mu = \nu (s-1). \label{lines13}
\end{eqnarray}
Solving this, we obtain
\begin{equation}
\epsilon(u) = -C u^{-(\mu+2)},
\end{equation}
where $C$ is a coefficient which is expressed in terms of $\nu$, $\mu$
and $s$.  In the rest of this subsection we will ignore coefficients
of order unity, such as $C$, since our primary interest is in the
scalings of various quantities.  Thus, we write the solution for
$T(u)$ as
\begin{equation}
T(u) \approx u^{-\mu}\left(1-{1\over u^2}\right). \label{Tapprox}
\end{equation}
Clearly, this solution gives the first two non-zero terms of a power
series expansion in $1/u$ of the paraboloidal solution (note that the
term proportional to $1/u$ vanishes), i.e., the first two nonvanishing
terms of an expansion around the point at infinity.  For many
purposes, the first term is good enough, but the second term is
essential for estimating the Lorentz factor of the force-free outflow,
as we now discuss.  Incidentally, to get an idea of the error we make
by neglecting coefficients of order unity, note that the power series
expansions of the two analytic solutions given in \S\S~3.1 and 3.2 are
given by
\begin{eqnarray}
T(u) &\approx& {1\over2u} \left(1-{1\over4u^2}\right),
\qquad \nu=1, \quad \mu=1, \\
T(u) &\approx& {1\over2u^2} \left(1-{3\over4u^2}\right), \qquad \nu=0,
\quad \mu=2.
\end{eqnarray}
The factors $1/2$, $1/4$ and $3/4$ are omitted when we write the
series expansion approximately as in (\ref{Tapprox}),

The stream function $P(R,z)$ corresponding to the above solution is
\begin{equation}
P(R,z) = R^\nu T(u) \approx {R^{\mu+\nu} \over z^\mu}
\left(1-{R^2\over z^2}\right).
\end{equation}
Since $P$ is constant on field lines, we can solve for the shape of a
field line with footpoint at $\Rfp$:
\begin{eqnarray}
{R\over\Rfp} &\approx& \left({z\over\Rfp}\right)^{\mu/(\mu+\nu)}
\left[1 + \left({\Rfp\over z}\right)^{2\nu/(\mu+\nu)}\right]
\nonumber\\
&=& \left({z\over\Rfp}\right)^{(s-1)/s} \left[1 + \left({\Rfp\over
z}\right)^{2/s}\right], \label{Rexpr}
\end{eqnarray}
where, in the final expression, we have used equation (\ref{mueq}) to
substitute for $\mu$.  From equations (\ref{BR}), (\ref{Bphi}),
(\ref{Bz}), we can estimate the three components of the magnetic
field:
\begin{eqnarray}
B_R(R,z) &\approx& {R^{\mu+\nu-1} \over z^{\mu+1}}, \\
B_\phi(R,z) &\approx& -M{R^{(\nu^2-2\nu+\mu\nu-\mu)/\nu}
\over z^{\mu(\nu-1)/\nu}} \left(1-{R^2\over z^2}\right), \\
B_z(R,z) &\approx& {R^{\mu+\nu-2} \over z^\mu} \left(1-{R^2\over
z^2}\right).
\end{eqnarray}
We have written the first two terms for $B_\phi$ and $B_z$, but only
the leading term for $B_R$ since this component of the field is very
weak at large $u$.

Consider now the variation of the field components as a function of
$z$ {\it along a given field line}.  To calculate this, we substitute
for $R$ from equation (\ref{Rexpr}):
\begin{eqnarray}
B_R\left({z\over\Rfp}\right) &\approx& \left({\Rfp \over z}\right)^{(2s-1)/s}, \\
B_\phi\left({z\over\Rfp}\right) &\approx& -M \left({\Rfp\over
z}\right)^{(s-1)/s}
\left[1-\left({\Rfp\over z}\right)^{2/s}\right] , \\
B_z\left({z\over\Rfp}\right) &\approx& \left({\Rfp\over
z}\right)^{2(s-1)/s} \left[1-\left({\Rfp\over z}\right)^{2/s}\right] .
\end{eqnarray}
From these, we obtain the following expressions for the poloidal and
total field strength along a field line:
\begin{eqnarray}
B_p^2\left({z\over\Rfp}\right) &\approx& \left({\Rfp\over
z}\right)^{4(s-1)/s}
\left[1-\left({\Rfp\over z}\right)^{2/s}\right], \\
B^2\left({z\over\Rfp}\right) &\approx& M^2 \left({\Rfp\over
z}\right)^{2(s-1)/s}\\
&\times& \left[1-\left({\Rfp\over z}\right)^{2/s} + {1\over M^2}
\left({\Rfp\over z}\right)^{2(s-1)/s}\right] \nonumber.
\end{eqnarray}
Substituting these expressions and (\ref{Rexpr}) in equation
(\ref{vmin}), we obtain the following estimate for the minimum Lorentz
factor along a field line,
\begin{equation}
\gamma_{\rm min} \approx \left[ \left({\Rfp\over z}\right)^{2/s} +
{1\over M^2}\left({\Rfp\over z}\right)^{2(s-1)/s} \right]^{-1/2}.
\label{gammaest}
\end{equation}
Since there are two terms, we need to keep track of the behaviour of
each.

First, consider the case $\nu>1$.  In this case, we saw earlier that
$1/s = \nu/2 > 1/2$.  Therefore, for large $z/\Rfp$, the second term
in equation (\ref{gammaest}) always dominates, and so we have
\begin{equation}
\gamma_{\rm min} \approx M \left({z\over\Rfp}\right)^{(s-1)/s} = M
\left({z\over\Rfp}\right)^{(2-\nu)/2}, \quad \nu > 1,
\end{equation}
where we have substituted for $s$ using equation (\ref{special}). For
$\nu=5/4$, this gives $\gamma_{\rm min} \propto z^{3/8}$, which agrees
very well with the scaling we find for the numerical solution shown in
Figure 4 (thick solid line).

Consider next $\nu < 1$.  In this case, the second term in equation
(\ref{gammaest}) dominates for a range of $z/\Rfp$ up to a certain
limit and the first term dominates beyond that, i.e.,
\begin{eqnarray}
\gamma_{\rm min} &\approx& M \left({z\over\Rfp}\right)^{(s-1)/s},
\qquad \nu < 1, \qquad z < z_{\rm crit}, \\
\gamma_{\rm min} &\approx& \left({z\over\Rfp}\right)^{1/s}, \qquad \nu
< 1, \qquad z > z_{\rm crit},
\end{eqnarray}
where the transition value of $z$ is given by
\begin{equation}
z_{\rm crit} = \Rfp M^{-s/(s-2)}.
\end{equation}
The two regimes of $\gamma_{\rm min}$ are seen clearly in the bottom
right panel of Figure 9.

Before concluding, we remind the reader once again that all the
relations given in this subsection have the correct scalings, but we
have ignored numerical coefficients.  For instance, the coefficient
$M$ in the expressions for $B_\phi$ and $\gamma_{\rm min}$ is perhaps
more appropriately written as $sM$.  However, $s$ is a quantity of
order unity and so, in the general spirit of neglecting coefficients,
we have ignored this refinement.  The other point is that we ignored
line 2 of the differential equation (\ref{Teq2}) in the discussion
here, whereas the terms in this line are negligible only for a limited
range of $s$.  A more complete analysis would retain these terms, but
this is beyond the scope of the paper.

\label{lastpage}

\end{document}